\documentclass[10pt,journal,comsoc]{IEEEtran}

\usepackage{cite}
\usepackage{times}

\usepackage{latexsym}
\usepackage{amsfonts}
\usepackage{paralist}
\usepackage{setspace}
\usepackage{color}

\usepackage{graphicx, url, xspace}
\usepackage{epstopdf}
\usepackage{subfigure}
\usepackage{caption}
\usepackage{indentfirst}
\usepackage{amsmath}
\usepackage[noend]{algorithmic}
\usepackage{algorithm}
\usepackage{amssymb}
\usepackage{multirow}
\usepackage[scaled=0.85]{beramono}

\newtheorem{theorem}{Theorem}

\newtheorem{definition}{Definition}

\makeatletter

\newcommand{\Rmnum}[1]{\expandafter\@slowromancap\romannumeral #1@}

\makeatother

\allowdisplaybreaks 
\def\ie{\textit{i.e.}\xspace}
\def\etal{\textit{et al.}\xspace}

\def\eg{\textit{e.g.}\xspace}

\def\st{\xspace\textit{s.t.}\xspace}

\newcommand{\CUTXY}[1]{{}}

\newcommand{\jnote}[1]{{$\langle${\textbf{\textcolor{blue}{#1}}}$\rangle$}}

\newcommand{\sn}{social network\xspace}
\newcommand{\dea}{de-anonymize\xspace}
\newcommand{\da}{de-anonymization\xspace}
\newcommand{\Da}{De-anonymization\xspace}
\newcommand{\gnp}{$\mathcal{G}(n,p)$\xspace}

\newcommand{\tp}{transition phenomenon\xspace}

\newcommand{\pk}{prior knowledge\xspace}
\newcommand{\ak}{adversarial knowledge\xspace}
\newcommand{\bk}{background knowledge\xspace}
\newcommand{\kg}{knowledge\xspace}

\newcommand{\sm}{subgraph matching\xspace}

\newcommand{\sgc}{subgraph counting\xspace}
\newcommand{\att}{attribute\xspace}
\newcommand{\atts}{attributes\xspace}
\newcommand{\pr}{probability\xspace}
\newcommand{\prs}{probabilities\xspace}
\newcommand{\pri}[1]{type-#1 privacy\xspace}

\renewcommand{\paragraph}[1]{\smallskip \noindent {\textbf{#1}}}
\renewcommand{\baselinestretch}{0.98}

\graphicspath{{./figures/}{./../figures/}{../authors/}}
\begin{document}


\title{Social Network De-anonymization: More Adversarial Knowledge, More Users Re-Identified?}

\author{Jianwei~Qian,
	Xiang-Yang~Li,
	Yu~Wang,
	Shaojie~Tang,
	Taeho~Jung,
	Yang~Fan%
\thanks{J. Qian and T. Jung are with the Department of Computer Science, Illinois Institute of Technology, USA.}%
\thanks{X.Y. Li and Y. Fan are with the School of Computer Science and Technology, University of Science and Technology of China, China.}%
\thanks{Y. Wang is with the School of Information and Communication Engineering, University of North Carolina at Charlotte, USA.}%
\thanks{S. Tang is with the Department of Information Systems, Naveen Jindal School of Management, the University of Texas at Dallas, USA.}%
}

\maketitle


\begin{abstract}
Following the trend of data trading and data publishing, many online social networks
have enabled potentially sensitive data to be exchanged or shared on the web.
As a result, users' privacy could be exposed to malicious third parties since they are extremely 
vulnerable to \da attacks, \ie, the attacker
links the anonymous nodes in the social network to their real identities with the help of \bk.
Previous work in social network \da mostly focuses on designing accurate and efficient \da methods.
We study this topic from a different perspective and attempt to  
investigate the intrinsic relation between the attacker's knowledge and the expected \da gain.
One common intuition is that the more auxiliary information the attacker has, the more accurate \da becomes. 
However, their relation is much more sophisticated than that.
To simplify the problem, we attempt to quantify \bk and \da gain under several assumptions.
Our theoretical analysis and simulations on synthetic and real network data
show that more background knowledge may \textit{not necessarily} lead to more \da gain in \textit{certain} cases.
Though our analysis is based on a few assumptions,
the findings still leave intriguing implications for the attacker to make better use of the \bk when performing \da, and for the data owners to better measure 
the privacy risk when releasing their data to third parties.
\end{abstract}

\begin{IEEEkeywords}
Social network, de-anonymization, background knowledge, adversarial knowledge, quantification.
\end{IEEEkeywords}

\section{Introduction}
\label{sec:introduction}

Nowadays we are able to collect and analyze data from various social networking sites like Facebook and Weibo, which may contain sensitive information about individuals. Typical social network data is stored in the form of graphs/networks, 
with nodes representing users and edges representing relations between uses. 
As social network analysis has great potential in many domains like business intelligence and social science, more and more open platforms
have emerged to enable social network data to be exchanged on the web \cite{gross2005information,chirag2014collaborative,xu2014identifying}.
However, users' privacy could be exposed to malicious attackers. 
Although users' IDs are usually removed before the data publishing, the attacker still can link the nodes in the graph to
their identities in real life by utilizing auxiliary information (a.k.a. background/adversarial knowledge), which is known as the \da attack \cite{narayanan2009anonymizing,backstrom2007wherefore}.

Social network de-anonymization has recently attracted wide attention from the social network research community. 
Most of prior work focuses on designing graph mapping techniques to de-anonymize the entire network
\cite{backstrom2007wherefore,narayanan2009anonymizing, 
ji2014structural, 
ji2015your}.
Their approaches are usually comprised of two phases: landmark identification and mapping propagation.
In addition, some researchers concentrated on de-anonymizing a small group of users, like ego networks, and proposed anonymization methods to defend against such attacks \cite{cheng2010k,zou2009k,zhou2008preserving,wang2013outsourcing,hay2007anonymizing}.  
It has been shown that most existing network data are de-anonymizable partially or even completely,
and still no existing countermeasures can effectively prevent such attacks \cite{ji2014structural}.

This paper provides a different perspective on social network de-anonymization. 
Our goal is to discover the internal relation between the attacker's background knowledge 
and her \textit{\da gain} (\ie the amount of user identity loss).
To this end, we are facing three major \textbf{challenges}.
\textit{Firstly,} it is extremely hard to quantify the \da gain of an attacker without actually performing a real \da attack. 
To study the influence of \bk on the \da gain, we must consider various scenarios to make the results as general as possible.
However, it is impossible to actually perform all the possible real attacks to count the users de-anonymized in all these cases.
Thus we must estimate the expected \da gain in a general way.
Given an imaginary attacker with a certain amount of \bk, it is still too complicated to estimate how many users would be de-anonymized,
which depends on what type of knowledge the attacker has and how she performs \da. All these obscure factors contribute to the difficulty of 
the quantification of \da gain.
\textit{Secondly,} it is very challenging to quantify \ak under a unified framework due to its large diversity. 
An attacker's \bk may include a variety of profiles (like gender, age, location)
and topological information such as node degrees (friend numbers), 1-hop neighborhood (ego network) and communities (interest groups).
\textit{Lastly,} the attacker capability is unpredictable.  For example, an attacker may gather auxiliary information from multiple sources, which may not be fully revealed to the public. She may design various delicate techniques to \dea the social network and 
her computation ability is also unknown to us.
To overcome these challenges, we will have to make several assumptions to simplify the problem.

In this paper, we model the \da attack on the basis of subgraph matching/isomorphism (but \da is much more complex than it). 
The released social network dataset is referred to as the \textit{published graph}, which is anonymous to the attacker. 
The attacker's background knowledge can also be modeled as a graph (referred to as the \textit{query graph}), in which she usually knows the users' identifications. The attacker aims to link the users in the query to the corresponding nodes in the published graph so as to acquire their private information. 
She maps users in the query graph to nodes in the published graph to find subgraphs that  
match the query. The definition of the verb ``match'' depends on the properties of \bk (Section \ref{sec:bk}).  
Previous works have proposed many approximation algorithms for efficiently searching for the subgraph(s) that match the query 
\cite{backstrom2007wherefore,narayanan2009anonymizing,ji2014structural}, 
so we assume the attacker is powerful enough to find all the possible subgraphs that match her \bk.
Since any of these subgraphs matches her \bk, she would be unable to tell them apart
and any of them is treated as a possible candidate of the real match of the query.
Now the difficulty of \da lies in the indistinguishability of the matched subgraphs. 
The fewer matched subgraphs are found, the easier it is to pinpoint the real match and the more information is gained by the attacker.
We will formally define \da gain in Section \ref{sec:pri-def} based on this intuition.
Meanwhile, we quantify the attacker's \bk from two aspects: by quantity (\eg node number of the query graph) 
and by quality (\eg the extent of ``particularity'' of nodes, edges and attributes) (Section~\ref{sec:bk}).
On top of that we will study the influence of \bk on \da gain from a fundamental view.

We present a detailed theoretical analysis for \gnp random graphs \cite{erdos1960evolution} 
and power-law graphs \cite{chung2002connected} (Sections~\ref{sec:ER} and \ref{sec:power}), and conduct rich simulations for 
both synthetic data and real network datasets (Section \ref{sec:experiment}).
The results show that in some settings, the attacker's \da gain is monotone increasing with the amount of \bk as expected by our intuition. 
However, it is not monotone increasing in other cases. For example, it could first decrease for a while, then go up after the
attacker's knowledge reaches a threshold (valley point), and finally  gets to the highest (vanish point)
(further explained in Sections~\ref{sec:ER} and \ref{sec:power}).
To clarify, we made several assumptions to simplify the problem because it is extremely challenging to directly analyze it.
Our findings are based on these assumptions so they only apply to certain situations. Yet the \tp we found in the relation between \bk and \da gain still leaves interesting implications for both data publishers and attackers (Section \ref{subsec:implic}).

Our \textbf{contributions} can be summarized as follows.
\begin{itemize}
\item We build a taxonomy of \bk in Section \ref{sec:bk} and comprehensively analyze the impacts of different types of \bk on the process and the result of \da.
\item To the best of our knowledge, this paper is the first attempt to quantitatively study the influence of \bk on \da gain. We present our definition
of \da gain and quantification of \bk and then present a detailed theoretical analysis of their relation.
It is revealed that more
\ak does not always result in more \da gain in certain cases. 
We further explain the reasons and the meaning of the critical points in their relation curve.
\item We present rich simulations on both synthetic and real network data sets even though we are facing the challenge of approximating the NP-complete \sm problem. The experiment displays different kinds of relation between \bk and \da gain, which validates our claim in one sense.
\end{itemize}

\section{Preliminaries}
\label{sec:prelim}
We first introduce subgraph matching,  the essence of social network de-anonymization attack, and then
define the de-anonymization attack on top of that.
We will also briefly introduce two popular graph models: \gnp random graphs and power-law graphs, 
which are to be used to model the social network.

\subsection{Subgraph Matching}
\label{sec:subg-match}

\begin{figure}
  \centering
  \includegraphics[width=0.4\textwidth]{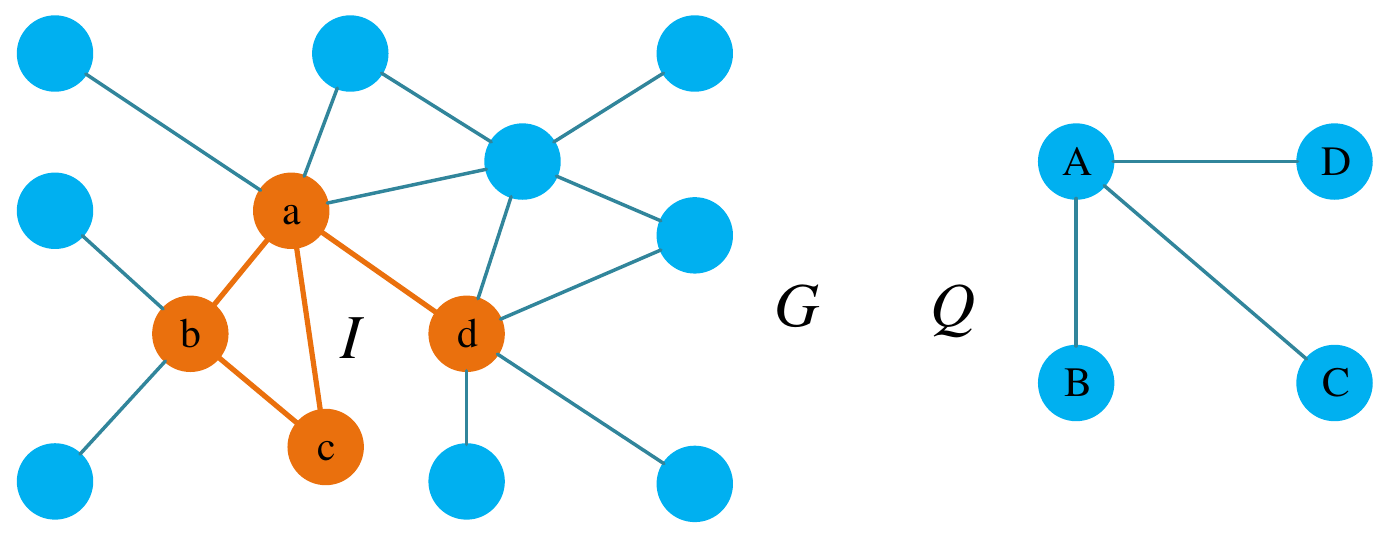} 
  \caption{\textbf{Subgraph matching:} $I$ is a randomly picked candidate of $Q$, \ie, an induced subgraph of $G$ yielded by randomly choosing $n_Q$ nodes in $G$ plus a mapping that maps them with nodes in $Q$. Whether $Q,I$ match is up to how we define match.}
  \label{subg-match}
\end{figure}

Subgraph matching (a.k.a. subgraph isomorphism) is a fundamental computational problem on graphs.
It aims to find a subgraph in a graph $G$ that is
isomorphic to another (usually smaller) graph $Q$.
We introduce the following important concepts that are critical to our model and analysis.
\begin{definition}[Subgraph]
Given a graph $G(V,E)$, a subgraph  $G_s(V_s,E_s)$ is a graph such that $V_s\subseteq V$ and $E_s\subseteq E$.
\end{definition}
\begin{definition}[Induced Subgraph]
Given a graph $G(V,E)$, an induced subgraph $I(V_{I},E_{I})$ is a graph \st $V_{I}\subseteq V$, and for any $u,v\in V_{I}$ with an edge in $E$, the edge exists in $E_{I}$. 
\end{definition}
\begin{definition}[Subgraph Matching]
Given a graph $G(V,E)$ and a query graph $Q=(V_Q,E_Q)$, \sm is the problem of finding a subset $V_s\subseteq V$ \st
there is a bijective mapping $f:V_q\rightarrow V_s$ that satisfies
$$\forall e(u,v)\in E_Q, e(f(u),f(v))\in E.$$
\end{definition}
The problem has been proved to be NP-complete \cite{cook1971complexity}. Fig.~\ref{subg-match} shows an example of subgraph matching.
The problem \textit{\sgc} is more related to this paper but even harder to solve, which requires counting the number of (or enumerating) 
matches (defined as follows) of $Q$ in $G$.
\begin{definition}[Match]
Given $G$ and $Q$, a match is a subset $V_s\subseteq V$ \textit{together with} a bijective mapping $f:V_Q\rightarrow V_s$ that satisfies
$$\forall e(u,v)\in E_Q, e(f(u),f(v))\in E.$$
\end{definition}
We denote the number of matches obtained from querying $G$ with $Q$ as $M(G,Q)$ ($M_Q$ for short) and the set of matches as $\mathcal{M}_Q$.

In this paper, we will analyze \da on the basis of subgraph matching. However, we study ``matching'' in a broad sense
since many real factors need to be taken into consideration, for example, the matching of nodal \atts, the induced and non-induced \sm,
exact and noisy matching, and probabilistic matching.

\subsection{Social Network De-anonymization}
\label{sn-da}
Social network \da \cite{backstrom2007wherefore,narayanan2009anonymizing} refers to the 
process of re-identifying anonymous nodes in a released social network.
The attacker usually possesses rich background \kg, \eg another \sn data set, so \da is a problem of mapping users in the \bk to nodes in the network.
We refer to the released \sn as the published graph $G$, and model the \bk as a query graph $Q$. Both $G,Q$ could contain rich nodal attributes.
\Da is similar to but much more than \sm.

\begin{definition}[De-Anonymization]
Given a published graph $G$, an attacker's de-anonymization on $G$ refers to an algorithm which, on inputs $G$ and $Q$, identifies a subset of nodes $V_I\subseteq V$ and a bijective mapping $f:V_Q\rightarrow V_I$ \st the subgraph $I$ induced in $G$ by $V_I$ \textit{matches} $Q$ under $f$ (denoted as $I\simeq^f Q$, or $I\simeq Q$ for simplicity).
Herein, the verb ``match'' can be defined in terms of structure/attribute similarity from multiple perspectives, which will be detailed in Section \ref{sec:bk}. 
Each pair of $V_I$ and the corresponding $f$ is referred to as a \textit{match} of $V_Q$.
There could be multiple matches (denote match number as $M_Q$) in $G$, but only one \textit{real match} of $Q$ exists whose nodes exactly correspond to real-life users in $Q$.
\end{definition}

To simplify, we assume $Q$ is connected (if not, the problem equals de-anonymizing multiple connected query graphs) and that the attacker de-anonymizes $Q$ as a whole.
The two major challenges of \da attack are finding all the subgraph matches ($\mathcal{M}_Q$) of a given query $Q$ and distinguishing the candidate matches.
Since \sm is NP-complete, researchers have been working on designing various approximation algorithms to overcome it, 
such as using landmark identification techniques to prune the search space \cite{backstrom2007wherefore,narayanan2009anonymizing,ji2014structural}.
So it is reasonable to assume this challenge can be overcome by a computationally powerful attacker, that is, she can find all the subgraphs of $G$ that match her \bk $Q$ by some means.
Thus, the real difficulty of \da lies in the indistinguishability of these matches since they all accord with $Q$, which introduces our definition of \da gain as below.

\subsection{De-Anonymization Gain} 
\label{sec:pri-def}
To quantify \da, we have to take the \bk $Q$ into account because it determines how much user identity information the attacker will recover.
As mentioned above, the attacker has used up all of her knowledge
to find the matches $\mathcal{M}_Q$ and every match exactly conforms to the attacker's \bk.
Thus, the attacker is unable to distinguish them, so she randomly picks a match $I\in\mathcal{M}_Q$ and treats it as the real correspond of $Q$. 
The \pr of picking the real match is  $1/M_Q$. 
Therefore, the quantity of matches reflects the \da gain in one sense, 
which is similar to $k$-anonymity \cite{sweeney2002k}.
The more matches are yielded by querying $G$ with $Q$, the less possible it is that user identities are recovered by the attacker. 
Let the number of nodes in $G,Q$ be $n,n_Q$ respectively. When the attacker picks the real match, $n_Q$ users are de-anonymized.
If she picks a wrong match, none (in the worst case) or part of the $n_Q$ users are de-anonymized. 
We cannot give a formula of the expected number of de-anonymized users but we know its infimum (\ie worst case) is $n_Q/M_Q$. 
We define \da gain as the expected fraction of de-anonymized users in the published graph $G$.
The infimum of \da gain is calculated as follows.
\begin{equation}
\label{equ:privacy}
DAG(Q)=\frac{n_Q}{M_Qn}.
\end{equation}
Obviously $DAG(Q)\in (0,1]$. When $M_Q=0$, the attacker could not find any match so we define $DAG(Q)=0$.
In the rest of this paper, when we mention \da gain we refer to its infimum.

\subsection{Graph Models}
\label{graph-model}

\begin{table}\small
\centering
 \begin{tabular} {| c | p{6cm} |}   
    \hline
     $G(V,E)$ & Published graph \\ \hline
     $Q(V_Q,E_Q)$ & Query graph \\ \hline
     \gnp & A random graph model with size $n$ and edge probability $p$ \\ \hline
     $n_Q$ & Size (vertex number) of $Q$ \\ \hline
     $m_Q$ & Edge number of $Q$ \\ \hline
     $ \beta$ &  Exponent of the power-law graph model \\ \hline
     $ M_Q$ & Short for $M(G,Q)$, the number of matches obtained from querying G with Q   \\ \hline
     $ I$ & Any candidate of $Q$, \ie, an induced subgraph of $G$ yielded by choosing $n_Q$ nodes in $G$, plus a mapping that maps them with nodes in $Q$  \\ \hline
     $DAG(Q)$ & Infimum of de-anonymization gain given query $Q$  \\ \hline
     $ Q\simeq I$ & $Q$ and $I$ are matched  \\ \hline
     $ Q\simeq_e I$ & $Q$ and $I$ are matched in edges \\ \hline
     $ Q\simeq_a I$ & $Q$ and $I$ are matched in node attributes \\ \hline
     $ p_{A}$ & Probability of two nodes sharing all the attribute \\ \hline
     $p(e_{ij})$ & The attacker's confidence on the edge $e_{ij}$ connecting nodes $i,j$ \\ \hline
 \end{tabular}
 \caption{Frequently used symbols}
\label{symb}
\end{table}

\subsubsection{\gnp Random Graph Model}
There are two closely related variants of the Erd\"{o}s-R\'{e}nyi (ER) model, but we focus on the \gnp model only.
In the \gnp model, a graph is constructed by connecting nodes randomly. 
Each edge is included in the graph with probability $p$ independent from every other edge.
As $p$ increases from 0 to 1, the graphs become denser with more edges.
Such random graphs are fundamental and useful for modeling problems in many
applications. However, a random graph in \gnp has the same expected degree
at every node and therefore does not capture some of the main behaviors of
numerous graphs developed in the real world.

\subsubsection{Power-Law Graph Model}
A graph is said to have power-law property if its degree sequences satisfy power-law distribution.
Namely, the fraction of nodes with degree $d$ is proportional to $d^{-\beta}$ for some constant $\beta$.
Many real world networks, at least asymptotically, conform to the power-law model 
for example, citation networks \cite{seglen1992skewness}  
 and social networks \cite{muchnik2013origins}.
Networks with power-law degree distributions are sometimes referred to as scale-free networks. 
Typically, the parameter $\beta$ for real world power-law networks is in the range $2<\beta<3$ \cite{choromanski2013scale}.
Power-law graphs can be generated with different methods, such as the preferential attachment mechanism 
(\eg Barab\'{a}si-Albert  model \cite{albert2002statistical}). 


\section{Attacker's Background Knowledge}
\label{sec:bk}
This section discusses about \bk of the attacker in terms of taxonomy and quantification, which covers most cases in real \da scenarios.

\subsection{A Taxonomy of Background Knowledge}
\begin{figure}[tbp]
  \centering
  \includegraphics[width=0.4\textwidth]{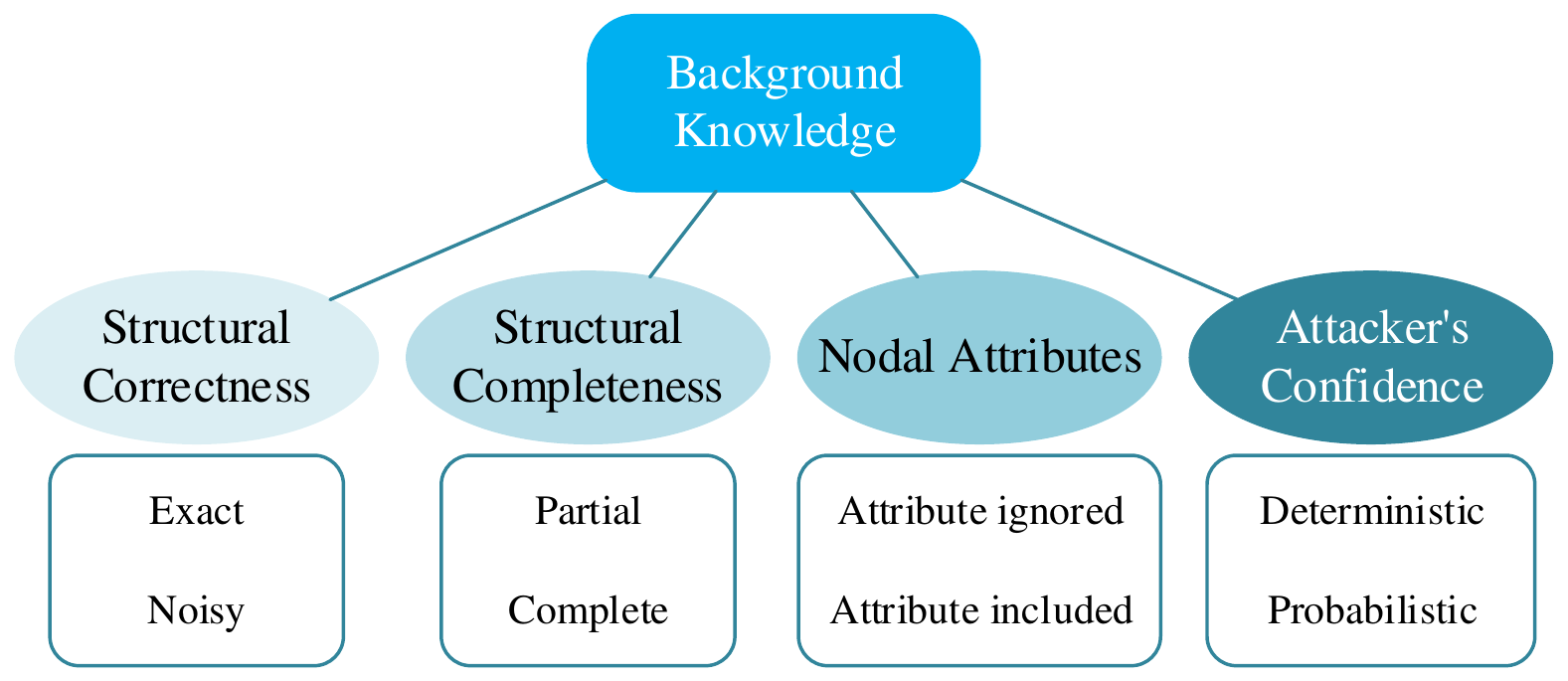} 
  \caption{\textbf{A taxonomy of background knowledge $Q$}}
  \label{bk-tax}
\end{figure}
The attacker' background knowledge in the de-anonymization helps her to re-identify the nodes in the published graph. Such knowledge 
may include the relationship between individuals (edges) and their attributes (nodal attributes). 
In this paper, background knowledge is represented with a query graph $Q$.
Different attackers may possess different types of background knowledge as their capabilities of information gathering and procession may vary. 
Such different knowledge has different impacts on the process of de-anonymization. Specifically, it influences how the attacker defines ``match''. 
As an example depicted in Fig.~\ref{subg-match}, given a \textit{candidate} $I$ of $Q$, that is, an induced subgraph yielded by randomly picking $n_Q$ nodes from $G$ and mapping them to the nodes in $Q$, 
the attacker determines if $I$ is a match of $Q$ on the basis of her beliefs in her knowledge $Q$.
Here we present a detailed taxonomy of the overall properties of background knowledge\footnote{Note that we do not classify \bk in terms of its sources or classes, which has been discussed in \cite{Qian2016} and has little impact on \da.}, which captures most, if not all, types of possible \da attacks with background knowledge in reality. 
We categorize the properties (the attacker's belief) of $Q$ in four dimensions: structural correctness, structural completeness, nodal attributes, and attacker's confidence. The former two lay emphasis on examining the occurrence of users' connections of $Q$ in $G$. An overview of the taxonomy is depicted in Fig.~\ref{bk-tax} and detailed explanations are as follows.

\textit{Structural correctness:} The structural information in $Q$, \ie edges, could be either exact or noisy.
\textit{Exact} means that the attacker believes every edge $e_{ij}$ in $Q$ is correct and indispensable; in other words,
she firmly believes that users represented by $i$ and $j$ are indeed connected in real life. 
Thus, a given subgraph of $G$ is considered to be a match of $Q$ only if it contains all the edges in $Q$.
In a more practical setting, structural information is often \textit{noisy}. The data publishers usually add noise/perturbation to the datasets before they are published in order to increase the difficulty of \da.
Such perturbation techniques include generalization, suppression, swapping and so on.
Besides, noise could be brought to $Q$ because of inaccurate information gathering.
In such circumstances, the attacker would assume her \kg about edges is noisy. 
When comparing their edges to determine whether a subgraph $I$ matches $Q$, the attacker would
tolerate a few missing or extra edges in $Q$.

\textit{Structural completeness:} The attacker's knowledge of structural information can be either partial or complete.
\textit{Partial} knowledge is caused by incomplete information gathering, which means incomplete background knowledge.
There might be a few edges missing in $Q$ that in fact exist in $G$. In this case, the attacker performs non-induced \sm as missing edges in $Q$ are allowed. 
Yet in a more ideal scenario, the attacker may think her \kg of edges is \textit{complete} and then perform induced subgraph matching.

\textit{Nodal attributes:} The query graph $Q$ might contain users' attributes (\textit{attribute included}) or not (\textit{attribute ignored}).
Sometimes attackers \dea graph data with only structural information as \pk, such as \cite{ji2014structural,ji2015your}.
In other cases, they also have the possession of users' profiles and utilize them in comparing users and the candidate nodes \cite{wang2013outsourcing,tai2011structural}.
The attribute included case can be further split according to the correctness of the attacker's knowledge
about \atts: the attacker 
1) performs exact matching on \atts if they are \textit{exactly correct}; or 2) allows a few errors if they are assumed to be \textit{almost correct}; or 3) matches the \atts of users and their candidates in an \textit{approximate} way, \eg, the ages of 32 and 34 might be treated as matched.

\textit{Attacker's confidence:} 
Most of previous works assume the attacker is very sure about her \kg of both edges and \atts, which is referred to as \textit{deterministic} \kg.
More practically, the attacker's knowledge could be \textit{probabilistic} since she could be uncertain about the information gathered. Details will be
presented in Section \ref{sec:probKg}.

\subsection{Background Knowledge Quantification}
Prior to the analysis of the influence of \bk on \da gain, we first present the quantification of \bk in two aspects.

\textit{By quantity:} The amount of \bk can simply be measured by its scalar quantity, such as node number, edge number, and  \att number.
We choose node number $n_Q$ as the metric for our theoretical analysis.

\textit{By quality:} The particularity of \bk also determines the success rate of \da. 
Special \kg , such as the ``outstanding'' nodes (a.k.a. seeds), edges, \atts, and patterns, can help the attacker greatly filter the candidate matches.
The more different the attacker's belief is from the global distribution, the more informative and valuable the knowledge is. 
Thus, \bk can be quantified by quality as well,
\eg, the number of nodes with high degrees, graph density, the number of special structural patterns like cliques, 
the ratio of edges whose \prs are close to 1 or 0 if the attacker has probabilistic \kg,
or the number of users with a particular \att whose distribution is very distinct from the global distribution (measured by Kullback-Leibler divergence).
We will also use the ratio of highly deterministic edges and 
graph density as the metrics. Both of them inflect the quality of the attacker's \bk on the structural information.

\section{Theoretical Analysis on \gnp Graphs}
\label{sec:ER}

Here we study the case where $G$ follows \gnp random graph model. For every type of \bk, we present formulas to calculate the match number $M_Q$, which decides the \da gain $DAG(Q)$ by Equation \ref{equ:privacy}. The formulas of $M_Q$ are derived from Theorem~\ref{theorem} below. 
We will plot the relation between $DAG(Q)$ and the amount of \bk 
in the end of this section.

\begin{theorem}
\label{theorem}
Given a graph $G\in \mathcal{G}(n,p)$ and any query graph $Q$ whose node number is $n_Q$ and edge number is $m_Q$, 
let $I$ be a randomly selected candidate of $Q$ (Fig.~\ref{subg-match}), 
the \textit{expected} number of matches of $Q$ in an ER random graph \gnp is
\begin{equation}
\mathbb{E}(M_Q)=\binom{n}{n_Q}n_Q!\cdot Pr(Q\simeq I),
\end{equation}
where $Q\simeq I$ means $Q$ and $I$ are matched.
We will omit the expectation symbol ($\mathbb{E}(\cdot)$) hereafter.
\end{theorem}
The proof is presented in Appendix \ref{app1}.
Notice this theorem does not make any assumption about $Q$.
It can be any \textit{given} graph with arbitrary structure, for instance an ego network or a 2-hop neighborhood graph. 

As mentioned in Section~\ref{sec:bk}, different types of \bk determine how match is defined.
For the sake of clarification, we first consider the case where $Q$ is attribute-ignored and deterministic,
and then study more complex scenarios on top of it.

\subsection{Deterministic Knowledge with Attribute Ignored}
Sometimes the attacker uses only structural information for \da,  
such as \cite{ji2014structural,ji2015your}.
To decide if $I$ is a match of $Q$, we only need to check if their edges match, so
$Pr(Q\simeq I)=Pr(Q\simeq_e I)$, where $Q\simeq_e I$ represents ``match in edges''.
There are four basic situations to be discussed.

\subsubsection{Exact \& Partial Knowledge}
\label{sec:exact-partial} 
At first, we consider the situation where the attacker thinks that her knowledge is exact but partial.
Then $I$ is a match if $Q$ is a subgraph of $I$, \ie, every edge in $Q$ has a counterpart in $I$ (the probability is $p$ because of the \gnp model).
Redundant edges in $I$ is acceptable and not of interest, thus we refer to it as \textit{subset matching}.
In this case, we have $$Pr(Q\simeq_eI)=p^{m_Q},$$
\begin{equation}
\label{equ:exact_partial}
M_Q=\binom{n}{n_Q}n_Q!\cdot p^{m_Q}.
\end{equation}
Note only node number and edge number of $Q$ are involved in this formula; its network structure can be arbitrary and has no influence on the result.

\subsubsection{Exact \& Complete Knowledge}
When the attacker thinks her background knowledge is exact and complete, she would perform \textit{exact matching} over the edges of $Q$ and $I$.
In this case, $I$ is not considered as a match if it has any redundant or missing edge. 
We have $$Pr(Q\simeq_eI)=p^{m_Q}(1-p)^{m_{0}-m_Q},$$
\begin{equation}
M_Q=\binom{n}{n_Q}n_Q!\cdot p^{m_Q}(1-p)^{m_{0}-m_Q},
\end{equation}
where $m_{0}=n_Q(n_Q-1)/2$ is the edge number of a complete graph with $n_Q$ nodes.

\subsubsection{Noisy \& Partial Knowledge}
This is a relaxation of the first case.
We use $\delta(Q,I)$ to denote the number of extra edges in $Q$ that do not exist in $I$.
Note this function $\delta$ is not symmetric.
The attacker performs matching, and $I$ is still considered to be a match of $Q$ if $\delta(Q,I)\leq\epsilon$
(at most $\epsilon$ unmatched edges).
Here $\epsilon$ is the threshold of acceptable error of the attacker's 
knowledge in terms of edges. 
In this scenario, 
$$Pr(Q\simeq_eI)=Pr(\delta(Q,I)\leq\epsilon)=\sum_{k\leq\epsilon}Pr(\delta(Q,I)= k),$$
\begin{equation}
M_Q=\binom{n}{n_Q}n_Q!\cdot \sum_{k\leq\epsilon}\binom{m_Q}{k}\cdot p^{m_Q-k}(1-p)^{k}.
\end{equation}

\subsubsection{Noisy \& Complete Knowledge}
In this case, a match of $Q$ in $G$ is ``almost'' an induced subgraph of $G$; there might be a few missing or extra edges.
To determine whether $I$ is a match, the requirement $\delta(Q,I)\leq\epsilon,\delta(I,Q)\leq\epsilon$ has to be satisfied.
In this case, $Pr(Q\simeq_eI)=Pr(\delta(Q,I)\leq\epsilon,\delta(I,Q)\leq\epsilon)=\sum_{k,l\leq\epsilon}Pr(\delta(Q,I)=k,\delta(I,Q)=l)$.
We can derive the formula of $M_Q$ similarly.

\subsection{Deterministic Knowledge with Attribute Included}
In the real world, users in a social network have attributes like gender, age and occupation. The attacker could also have 
such information in her background knowledge in addition to the
relations between users. 
Accordingly, each node in $G$ and $Q$ is attached with $n_A$ attributes, denoted as $A_i,0\leq i\leq n_A$.
Now $Q$ and $I$ are considered to be a match only if both of their edges and node attributes are considered as matched.
$Pr(Q\simeq I)=Pr(Q\simeq_e I)\cdot Pr(Q\simeq_a I).$
Likewise, $\simeq_a$ represents ``match in attributes''.

Each of the attributes has a domain, denoted as $A_i, 0\leq i\leq n_A$. Let $Pr_{A_i}(a)$ be the probability of a node being labeled as $a$, for any $a\in A_i$.
Then the expected probability of any two nodes sharing attribute $A_i$ is $p_{A_i}=\sum_{a\in A_i}Pr_{A_i}(a)Pr_{A_i}(a)$.
(An continuous attribute can be binned into buckets and thus converted to a discrete attribute.)
The expected probability of any two nodes sharing all the attributes is $p_{A}=\prod_{i=0}^{n_A}p_{A_i}$. 
Thus, we have $Pr(Q\simeq_a I)=p_A^{n_Q}$.

When the attacker has exact and partial background knowledge on edges (it can be easily extended to other cases), 
and exactly correct knowledge on attributes, 
the expected number of matches she will obtain by querying $G$ with $Q$ is
\begin{equation}
M_Q=\binom{n}{n_Q}n_Q!\cdot p^{m_Q}p_A^{n_Q}.
\end{equation}

\subsubsection{Almost Correct Knowledge on Attributes}
Similar to almost correct knowledge on edges, the attacker can have a few inaccurate information (caused by imperfect information gathering and data perturbation on $G$) 
about users' attributes
(at most $\epsilon$ pairs of nodes with mismatched attributes).
\begin{equation}
Pr(Q\simeq_a I)= \sum_{k\leq\epsilon}\binom{n_Q}{k}p_A^{n_Q-k}(1-p_A)^{k}.
\end{equation}

``Almost correct'' can also be interpreted in another way. Two users are still considered to match in attributes if they disagree in at 
most $\epsilon$ out of $n_A$ attributes ($\epsilon<n_A$). Then, $p_{A}$ should be modified to 
\begin{equation}
p_{A}=\sum_{\mathbb{S}\subseteq\mathbb{A},|\mathbb{S}|\leq\epsilon} \prod_{A_i\in\mathbb{S}}(1-p_{A_i}) \prod_{A_i\in\mathbb{A}/\mathbb{S}} p_{A_i},
\end{equation}
where $\mathbb{A}$ is the set of all the attributes, and $\mathbb{S}$ is the set of unmatched attributes.

\subsubsection{Approximate Attribute Matching}
Due to the noise added by the publisher to graph $G$, a few attribute values in $G$ might be twisted. 
A node corresponding to a real life person might have slightly different attribute values than those of the person.
For example, a person is 31 years old, but her age is distorted to 30 in $G$.
If the attacker is aware of the possible discrepancies between her background knowledge and the information presented by $G$,
it is very likely she ignores minor attribute discrepancies while performing subgraph  matching  attack.
To estimate the number of results yielded by subgraph query under this circumstance, we need to redefine $p_{A_i}$.
\begin{equation}
p_{A_i}=\sum_{a\in A_i}\sum_{b\in A_i}Pr_{A_i}(a)Pr_{A_i}(b)Pr(a\simeq b),
\end{equation}
where $Pr(a\simeq b)$ represents the probability that $a$ and $b$ approximately match.
As an example for age, the probability that 30 and 40 are an approximate match is 0, but the probability of 30 and 32 can be set to 1.
The probability is conceptually different from similarity functions, but it can be calculated using them such as cosine similarity, Dirac delta function or $exp(-x)$.  
We do not expand the formula of $Pr(a\simeq b)$ here because there can be different definitions for different attributes. 

\subsection{Probabilistic Knowledge}
\label{sec:probKg}
In a more realistic setting, the attacker's knowledge on users' relations and attributes could be probabilistic. 
Suppose a probability $p(e_{ij})$ is assigned 
to each edge $e_{ij}$ of the query graph $Q$ to represent the attacker's confidence over $e_{ij}$.
Now $Q$ is a complete graph and edges that do not exist is with zero confidence.
In this case, there is no need to distinguish exact or noisy knowledge.
$Q$ can be any one of the $2^{m_0}$ configurations generated from $V_Q$.
We can compare each configuration (denoted as $Q_c$) with $I$ to calculate the probabilistic that $Q$ and $I$ are matched.
\begin{equation}
\label{equ:probKg}
\begin{aligned}
& Pr(Q\simeq_e I) \\
= & \sum_{Q_c}Pr(Q_c|Q)\cdot Pr(Q_c\simeq_eI) \\
= & \sum_{Q_c}\prod_{e_{ij}\in E(Q_c)}p(e_{ij})\prod_{e_{ij}\notin E(Q_c)}(1-p(e_{ij}))\cdot Pr(Q_c\simeq_e I).
\end{aligned}
\end{equation}

\subsubsection{Probabilistic \& Partial Knowledge}
Due to the partial knowledge assumption, we have
$Pr(Q_c\simeq_e I)=p^{|E(Q_c)|}$.
The computation cost of $Pr(Q\simeq_e I)$ is exponential. One simplifying method is to assume $p(e_{ij})=p_e, \forall e_{ij}\in E_Q$.
Following Equation \ref{equ:probKg}, we can derive 
\begin{equation}
\begin{aligned}
& Pr(Q\simeq_e I) \\
= & \sum_{Q_c}Pr(Q_c|Q)\cdot Pr(Q_c\simeq_e I) \\
= & \sum_{k=0}^{m_0}\binom{m_0}{k}\cdot p_e^k(1-p_e)^{m_0-k}\cdot p^k \\
= & (1-p_e+pp_e)^{m_0}.
\end{aligned}
\end{equation}
An alternative simplification is to sample from the $2^{m_0}$ configuration and then rescale the result. 

\subsubsection{Probabilistic \& Complete Knowledge}
When the attacker believes her knowledge is complete,
$Pr(Q_c\simeq_e I)=p^{|E(Q_c)|}(1-p)^{m_0-|E(Q_c)|}$.
When it is assumed all $p(e_{ij})=p_e$,
we can derive 
\begin{equation}
\begin{aligned}
& Pr(Q\simeq_e I) \\
= & \sum_{k=0}^{m_0}\binom{m_0}{k}\cdot p_e^k(1-p_e)^{m_0-k}\cdot p^k(1-p)^{m_0-k} \\
= & (1-p-p_e+2pp_e)^{m_0}.
\end{aligned}
\end{equation}
A more realistic simplification is to assume that the attacker's confidence on edges of the complete 
graph $Q$ has three levels: high ($p(e_{ij})=p_1$ say 0.9), low ($p(e_{ij})=p_0$ say 0.1) and medium (say $p(e_{ij})= p$ for \gnp). 
For edges with medium confidence, the attacker has no additional knowledge of them, so she will overlook the checking of their 
occurrence in $I$.
Given $p_1,p_0$ and suppose the number of the three types of edges are $x_1,x_0,x_m$ respectively ($x_1+x_0+x_m=m_0$), then
\begin{equation}
Pr(Q\simeq_e I)=(pp_1+(1-p)(1-p_1))^{x_1}(pp_0+(1-p)(1-p_0))^{x_0}.
\end{equation}
We can measure \bk by the ratio of \textit{highly deterministic} edges $r=\frac{x_1+x_0}{m_0}$, which reflects the quality of the knowledge.
We will also plot the influence of $r$ on $DAG(Q)$.

\subsubsection{Probabilistic Attributes}
The attacker could also have uncertain knowledge about users' attributes. 
Suppose for each nodal attribute in $Q$, there is a probability distribution $p_i$ over the domain of the attribute reflecting the attacker's belief.
If she has no additional knowledge about a person's attribute $A_i$, her belief is set to the original distribution in $G$, that is $Pr_{A_i}$.
Then the probability of two nodes sharing attribute $A_i$ is
$p_{A_i}=\sum_{a\in A_i}Pr_{A_i}(a)\sum_{b\in A_i}p_i(b) Pr(a\simeq b)$,
where $Pr(a\simeq b)$ represents the probability that $a$ and $b$ match.

\subsection{Analytical Results}
\label{ER-analysis}
\begin{table}[!t]\small
\centering
 \begin{tabular}{ c | p{4.5cm} | c }   \hline
    Parameter & Meaning & Default \\ \hline \hline
     $n$ & size of the published graph $G$ & 1000,000 \\ \hline
     $p$ & edge probability in \gnp for $G$ & 0.2 \\ \hline
     $n_Q$ & size of query graph $Q$ & 50 \\  \hline
     $p_q$ & graph density of $Q$ & 0.3 \\ \hline
     $p_A$ & prob. of 2 nodes sharing attributes & 0.001 \\ \hline
     $p_e$ & attacker's confidence on any edge & 0.4  \\ \hline
     $p_1$ & attacker's high confidence on edges & 0.9  \\ \hline
     $p_0$ & attacker's low confidence on edges & 0.1  \\ \hline
     $r$ & ratio of highly deterministic edges & 0.5  \\ \hline
 \end{tabular}
 \caption{\textbf{Parameters used in plotting and default settings.} We set a default value for each of them in order to control variables when plotting each of their influence on the results. All possible parameter values are tried in addition to the default values.}
\label{tab:matlab-params}
\end{table}

\begin{figure*}[tbhp]
\centering
\subfigure[Exact \& partial: varying $p$]{\label{exact_partial_p}\includegraphics[width=0.32\linewidth]{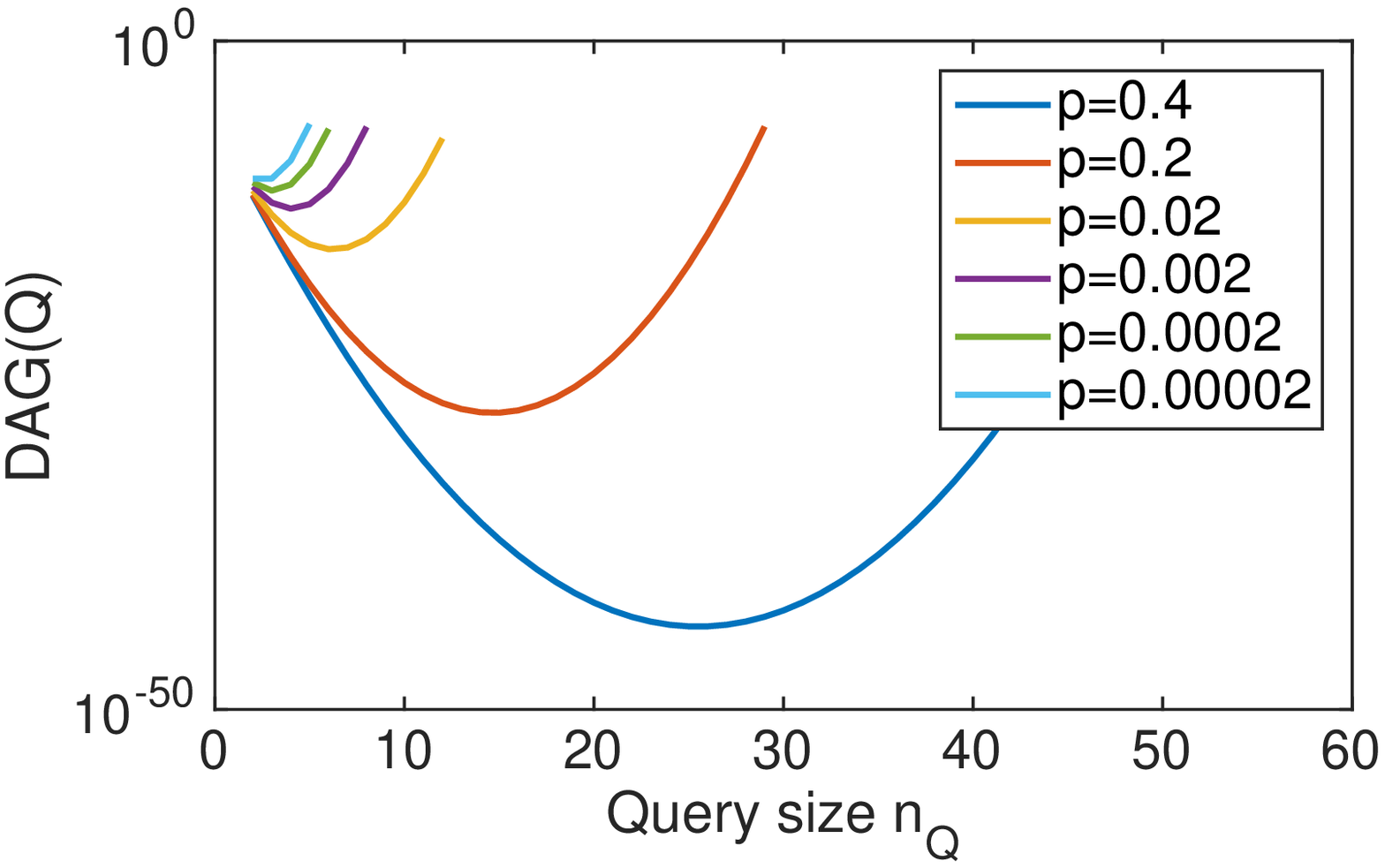}} 
\subfigure[Exact \& partial: varying $p_q$]{\label{exact_partial_pq}\includegraphics[width=0.32\linewidth]{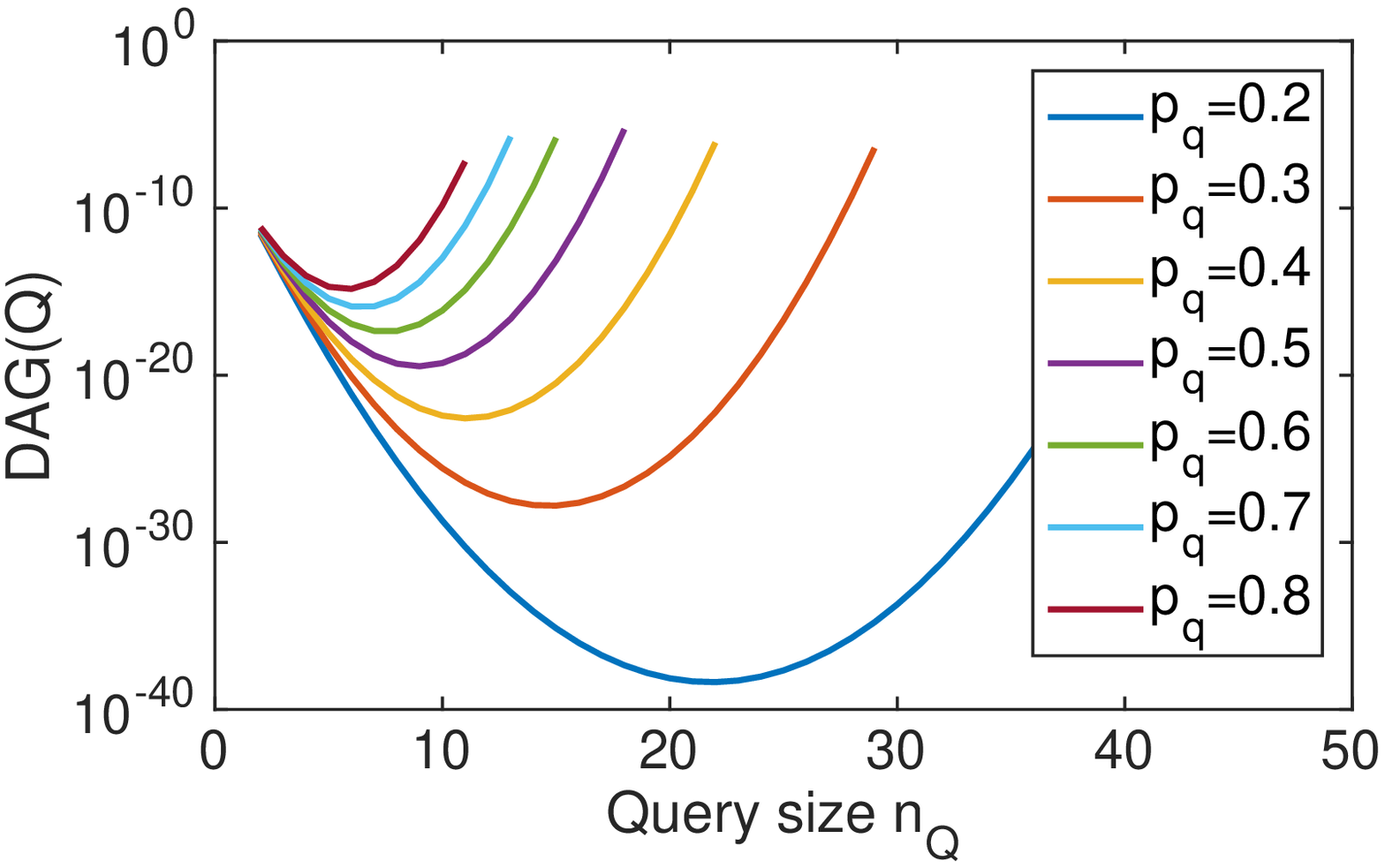}}
\subfigure[Exact \& partial: varying $p_A$]{\label{exact_partial_pA}\includegraphics[width=0.32\linewidth]{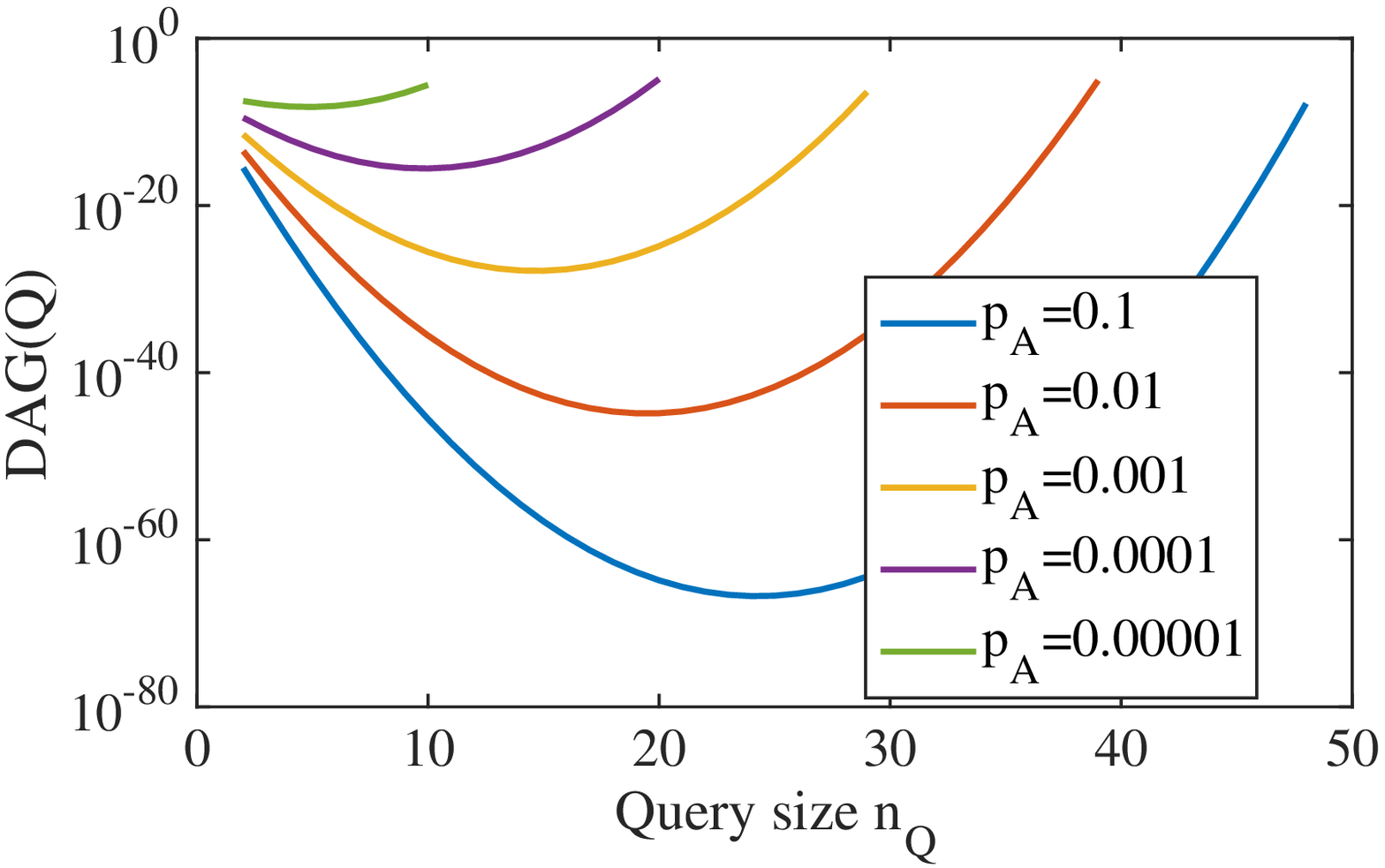}}\\
\subfigure[Exact \& complete: varying $p$]{\label{exact_complete_p}\includegraphics[width=0.32\linewidth]{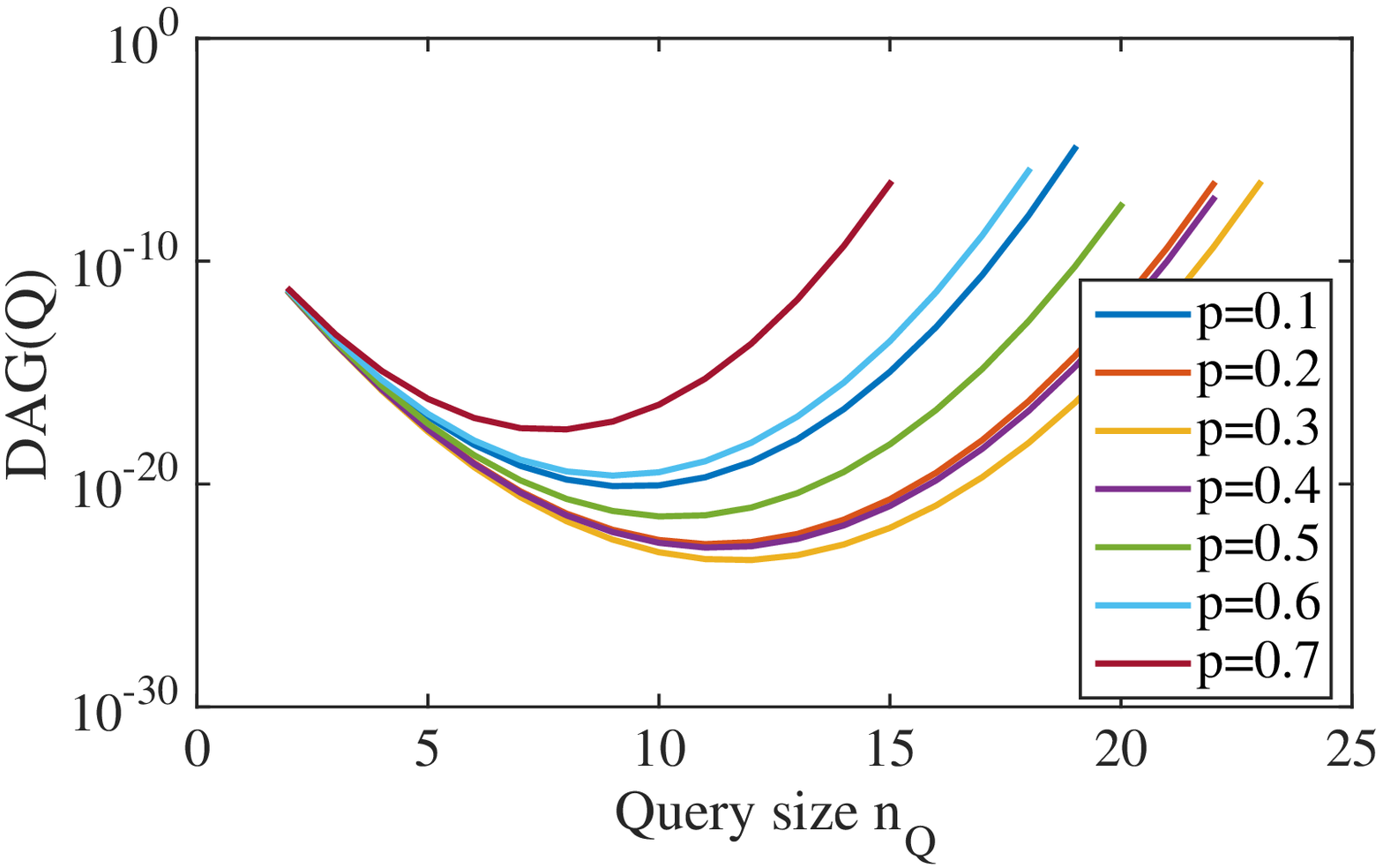}}
\subfigure[Exact \& complete: varying $p_q$]{\label{exact_complete_pq}\includegraphics[width=0.32\linewidth]{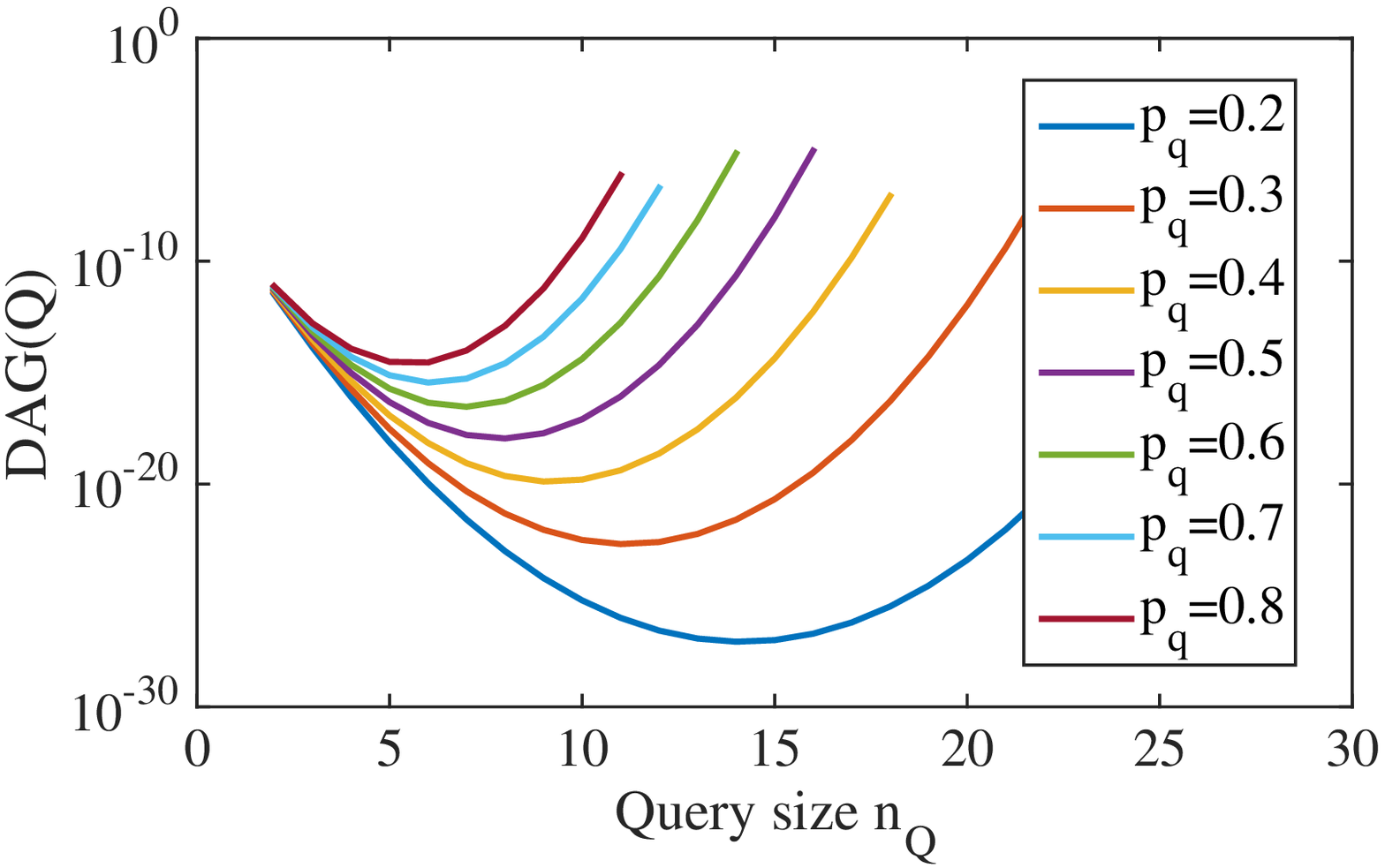}}
\subfigure[Exact \& complete: varying $p_A$]{\label{exact_complete_pA}\includegraphics[width=0.32\linewidth]{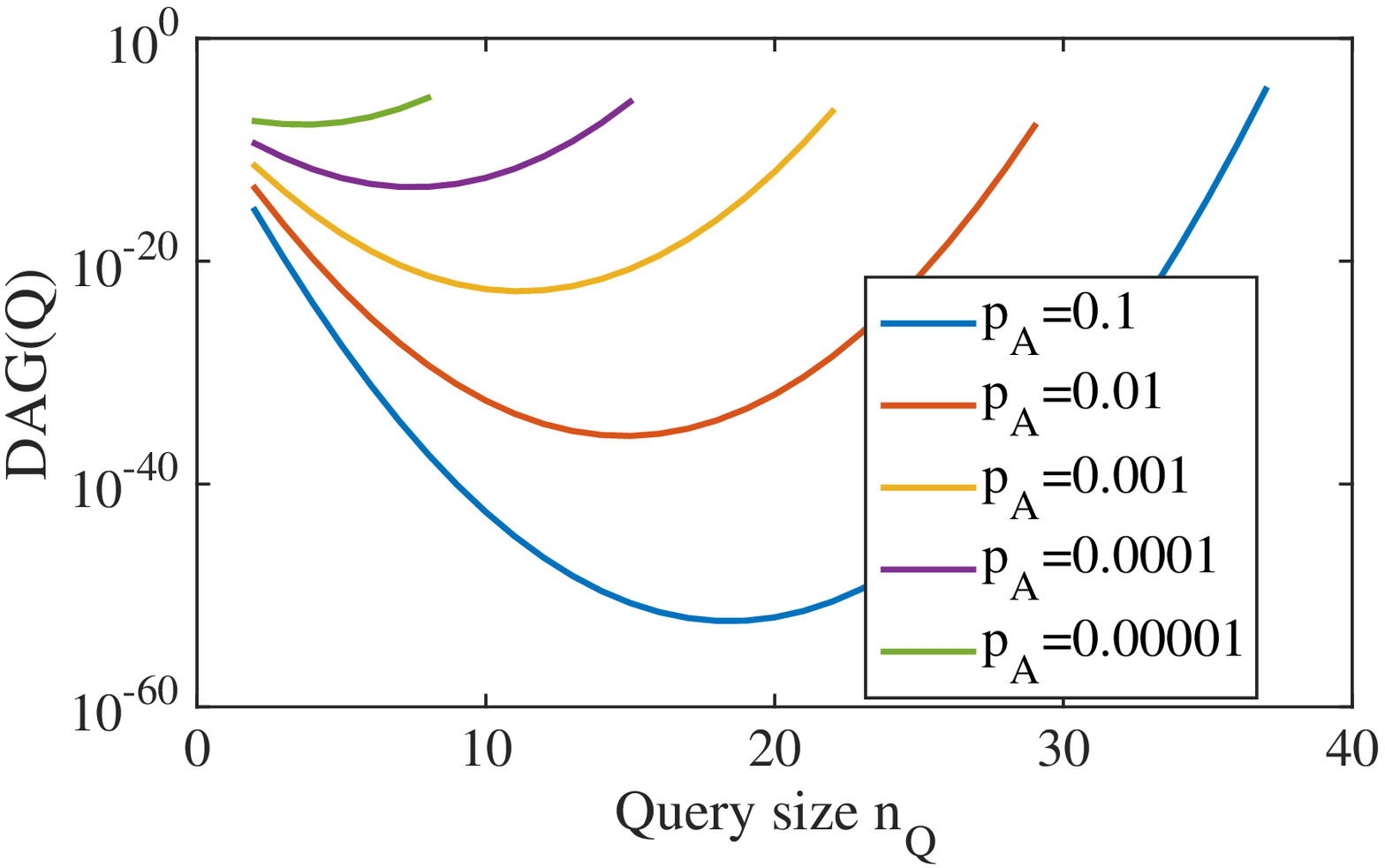}}
    \caption{\textbf{The impact of $n_Q$ on \da gain for exact \& partial(Fig. a-c)/complete(Fig. d-f) knowledge on \gnp graphs.} 
    Most curves have a transition phenomenon: with growing $n_Q$, \da gain first descends to the valley then rises to the maximum.}
    \label{exact_partial}\label{exact_complete}
\end{figure*}
\begin{figure*}[tbhp]
\centering
\subfigure[Noisy \& partial: varying $p$]{\label{noisy_partial_p}\includegraphics[width=0.24\linewidth]{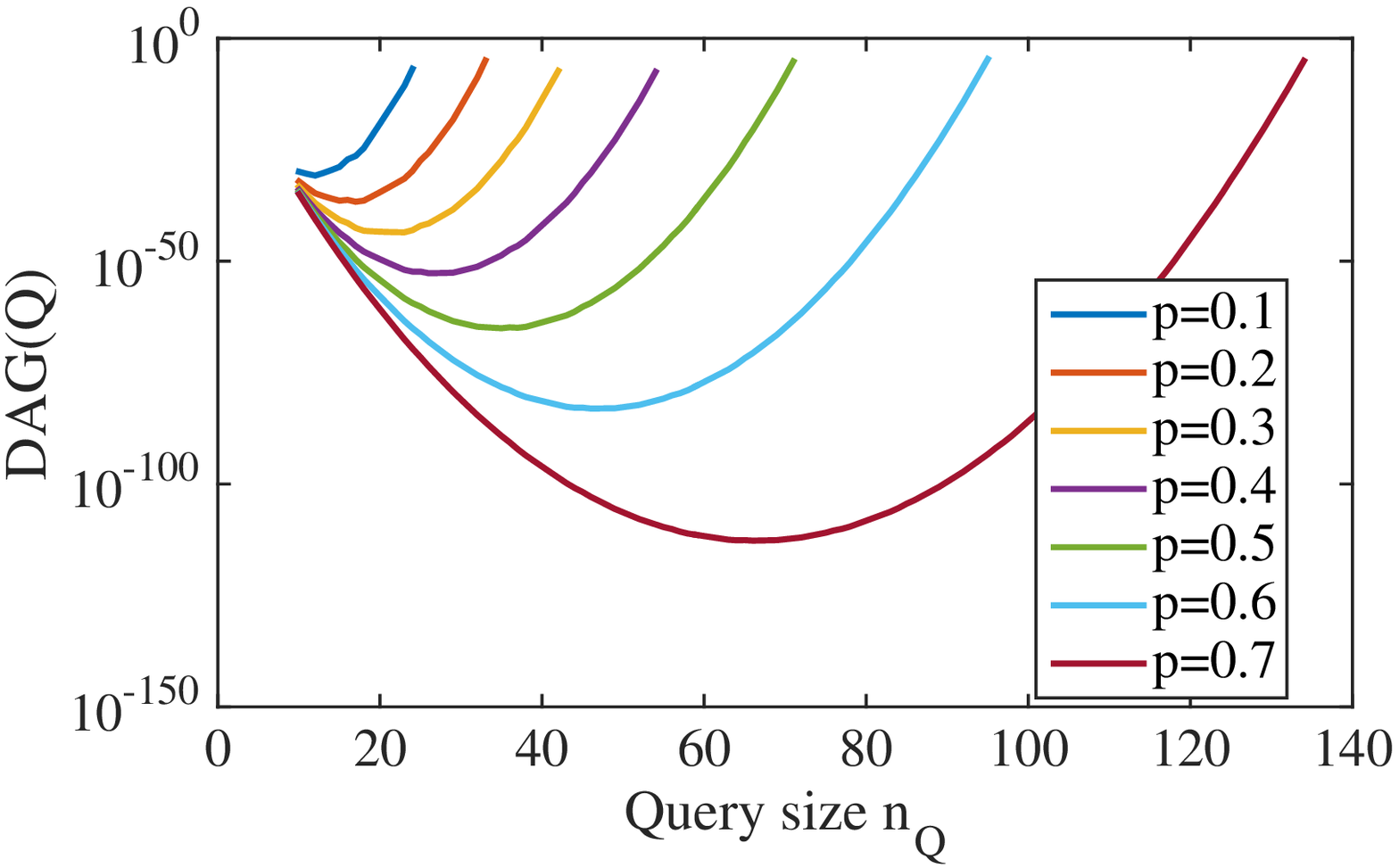}}
\subfigure[Noisy \& partial: varying $p_q$]{\label{noisy_partial_pq}\includegraphics[width=0.24\linewidth]{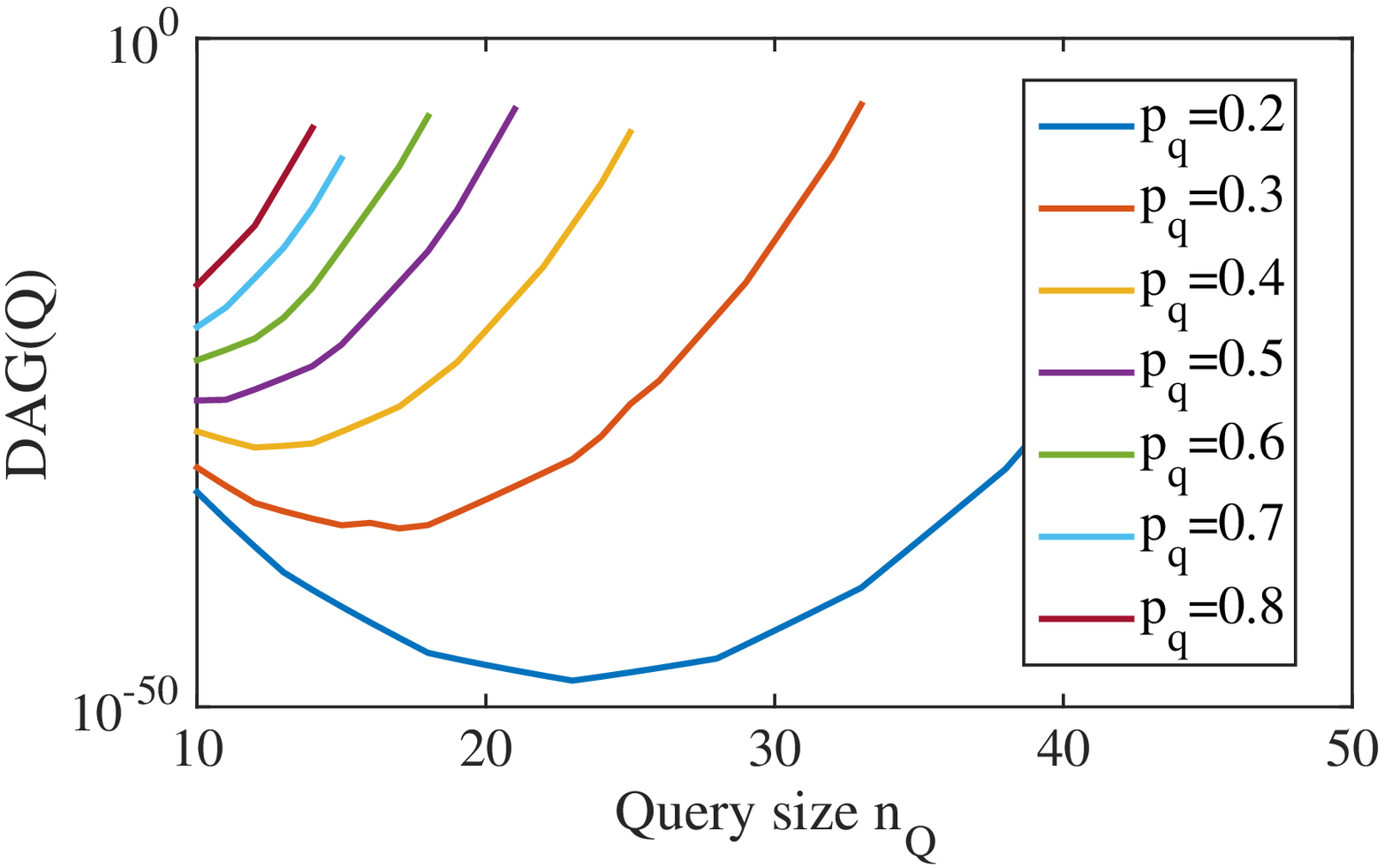}}
\subfigure[Noisy \& partial: varying $p_A$]{\label{noisy_partial_pA}\includegraphics[width=0.24\linewidth]{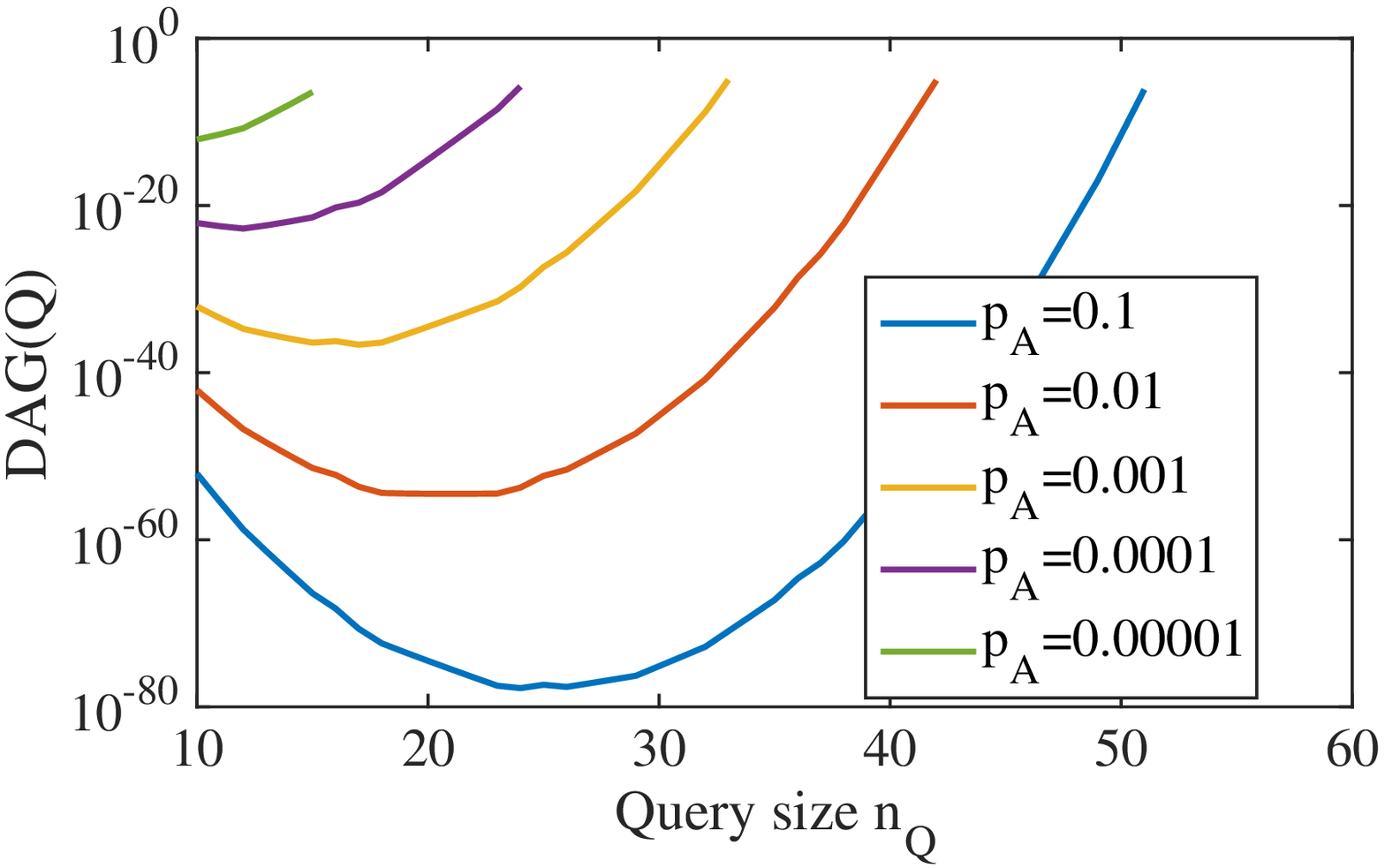}}
\subfigure[Noisy \& partial: varying $\epsilon$]{\label{noisy_partial_eps}\includegraphics[width=0.24\linewidth]{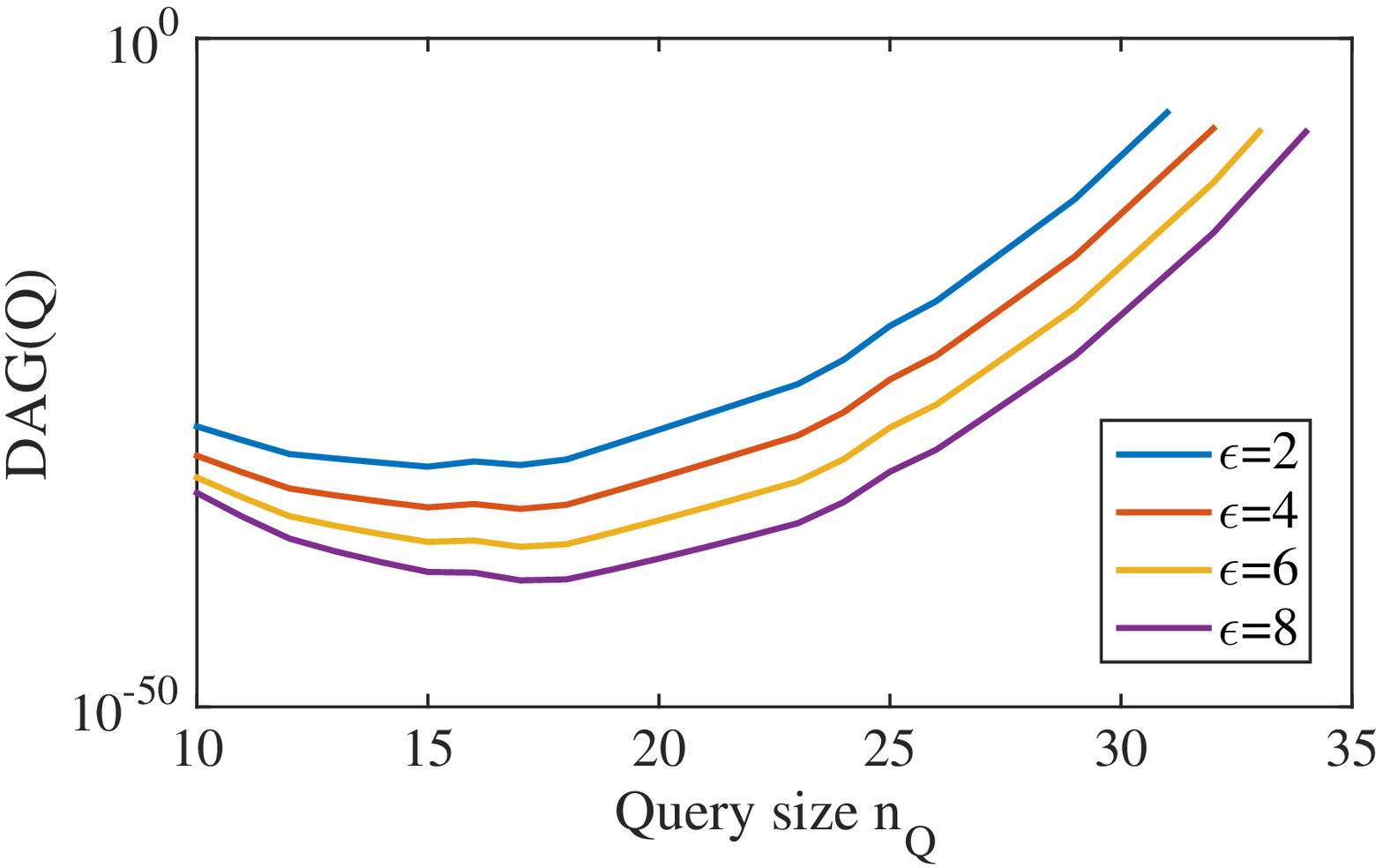}}\\
\subfigure[Noisy \& complete: varying $p$]{\label{noisy_complete_p}\includegraphics[width=0.24\linewidth]{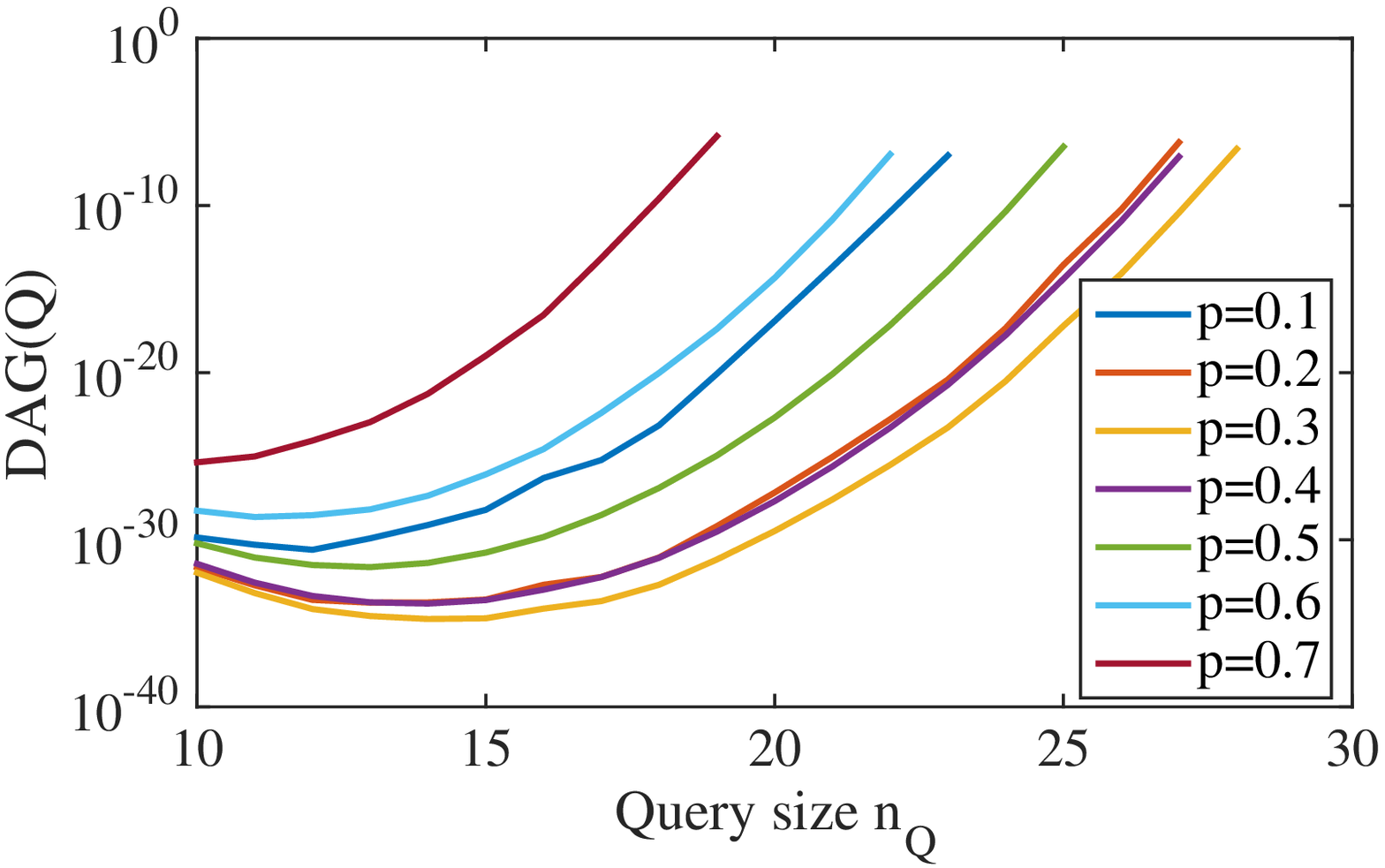}}
\subfigure[Noisy \& complete: varying $p_q$]{\label{noisy_complete_pq}\includegraphics[width=0.24\linewidth]{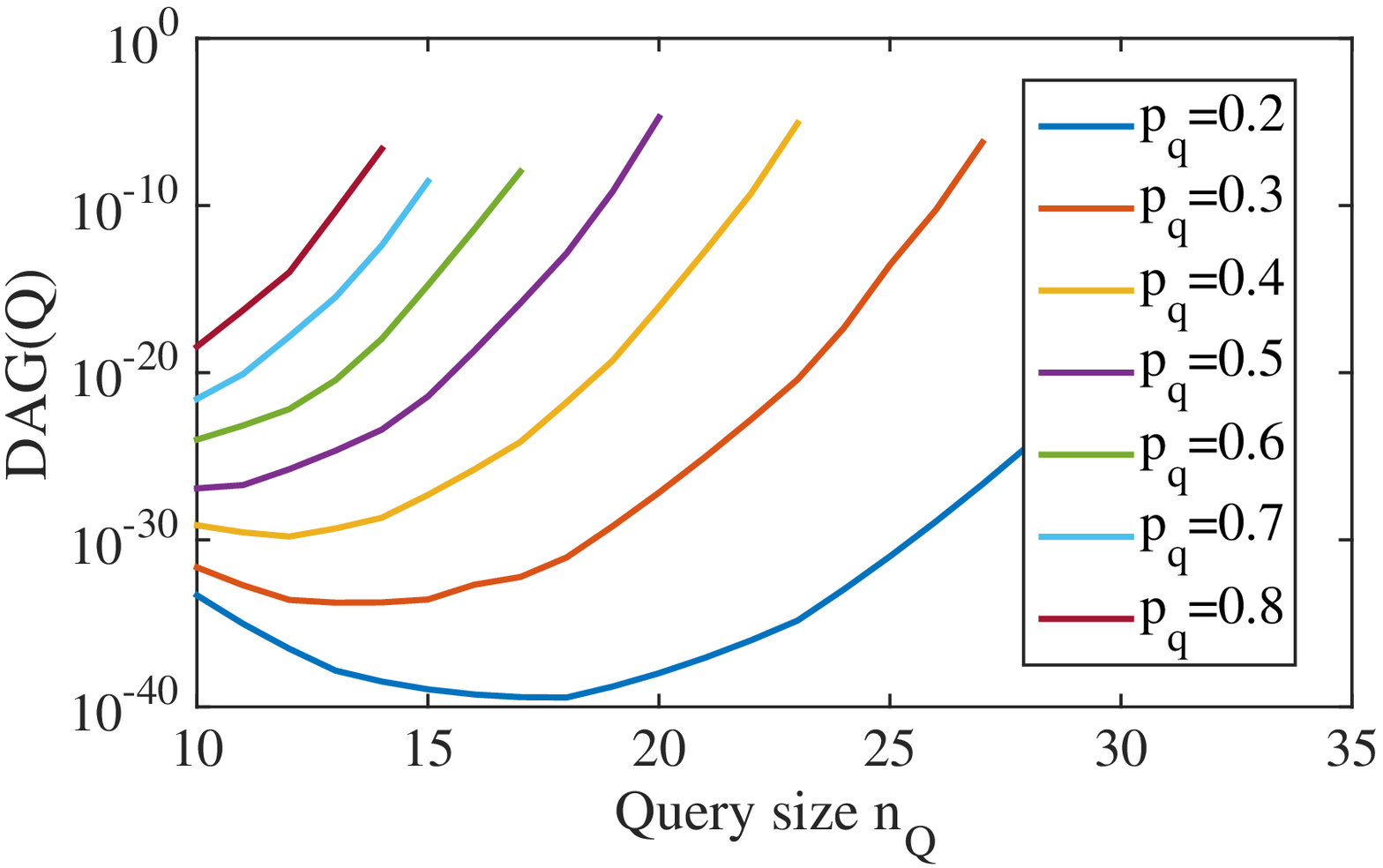}}
\subfigure[Noisy \& complete: varying $p_A$]{\label{noisy_complete_pA}\includegraphics[width=0.24\linewidth]{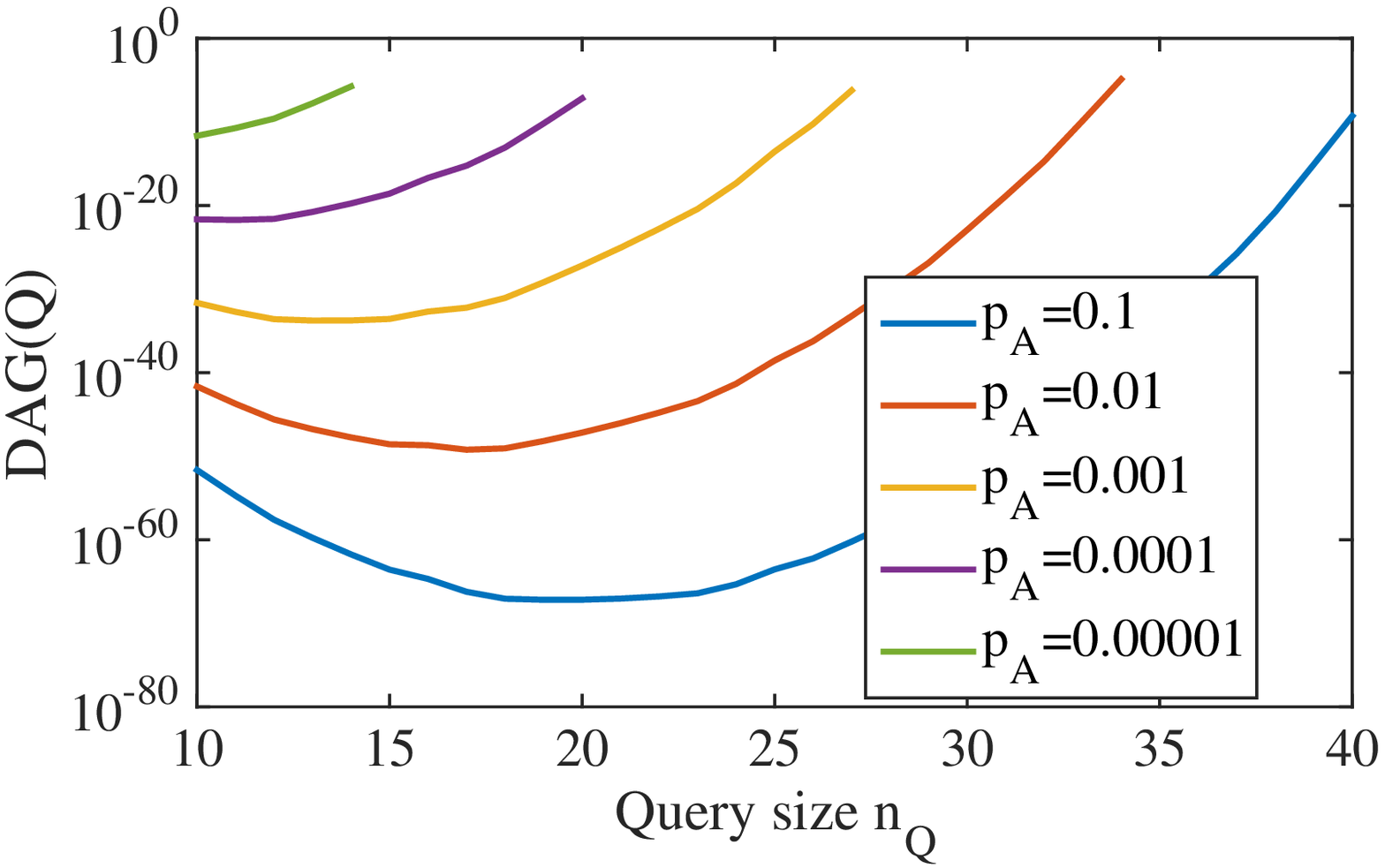}}
\subfigure[Noisy \& complete: varying $\epsilon$]{\label{noisy_complete_eps}\includegraphics[width=0.24\linewidth]{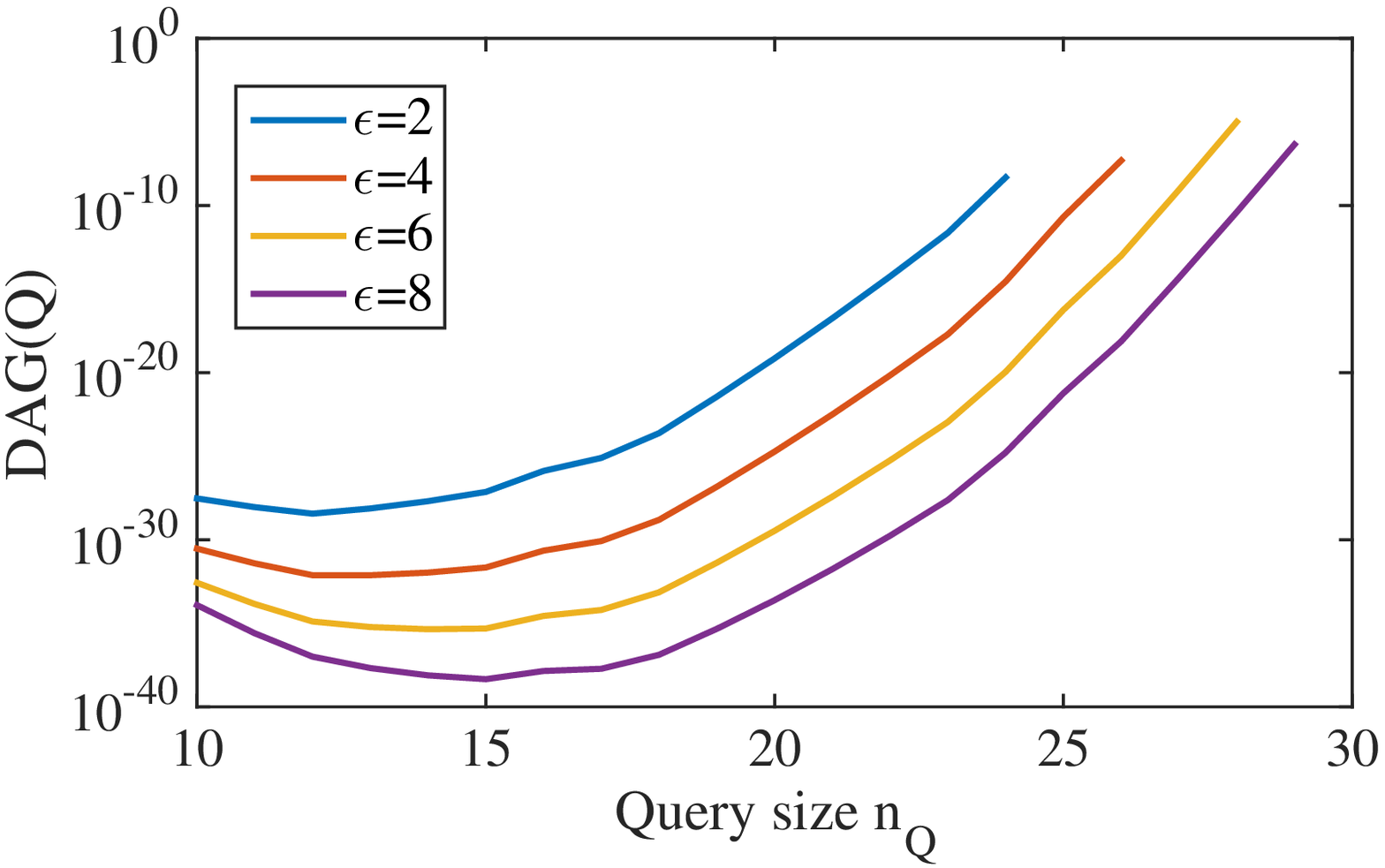}}
    \caption{The impact of $n_Q$ on \da gain for noisy \& partial(Fig. a-d)/complete(Fig. e-h) knowledge on \gnp.}\label{noisy_partial}\label{noisy_complete}
\end{figure*}
\begin{figure*}[tbhp]
\centering
\subfigure[Varying $p$]{\label{edgesPr_partial_p}\includegraphics[width=0.32\linewidth]{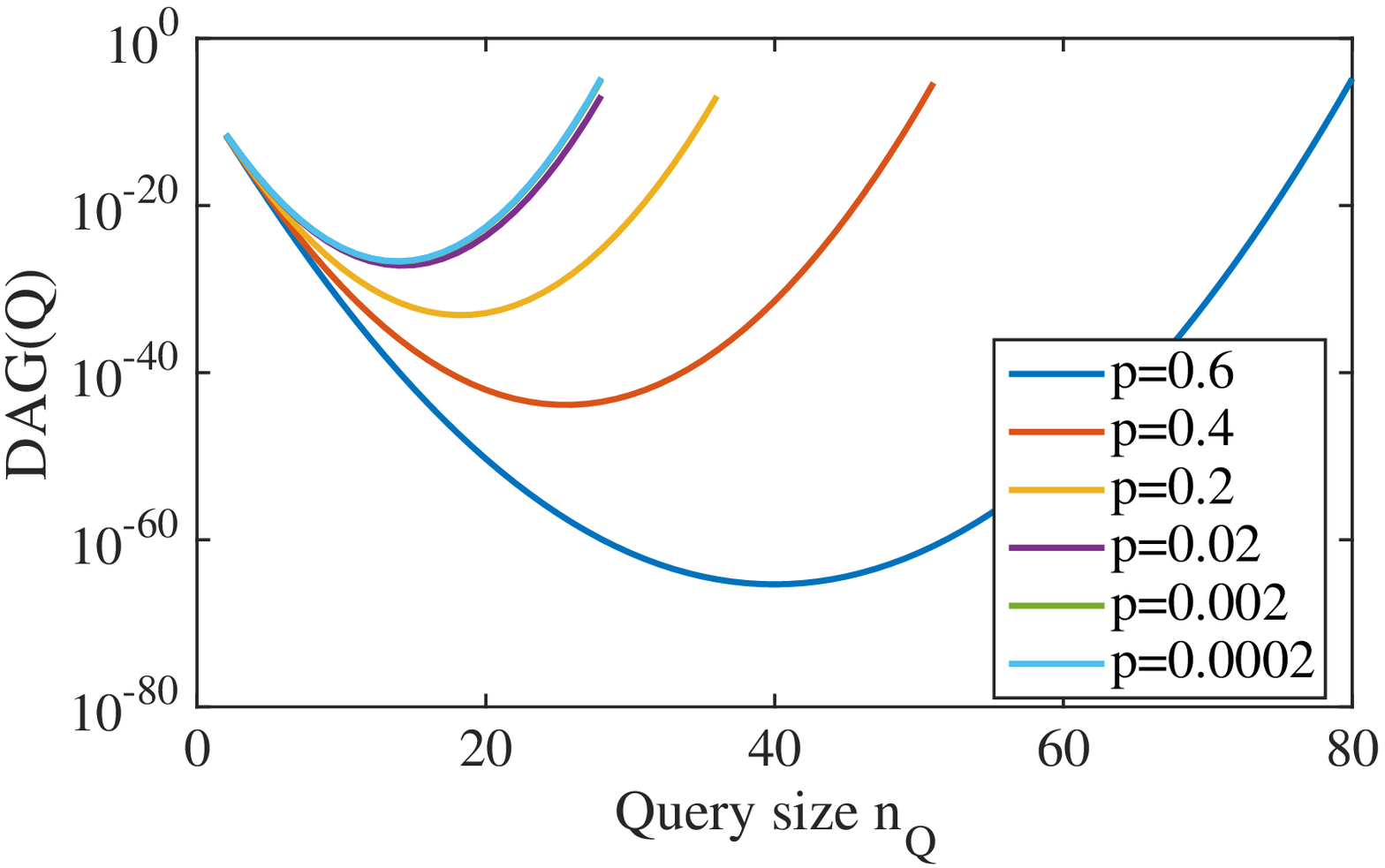}} 
\subfigure[Varying $p_A$]{\label{edgesPr_partial_pA}\includegraphics[width=0.32\linewidth]{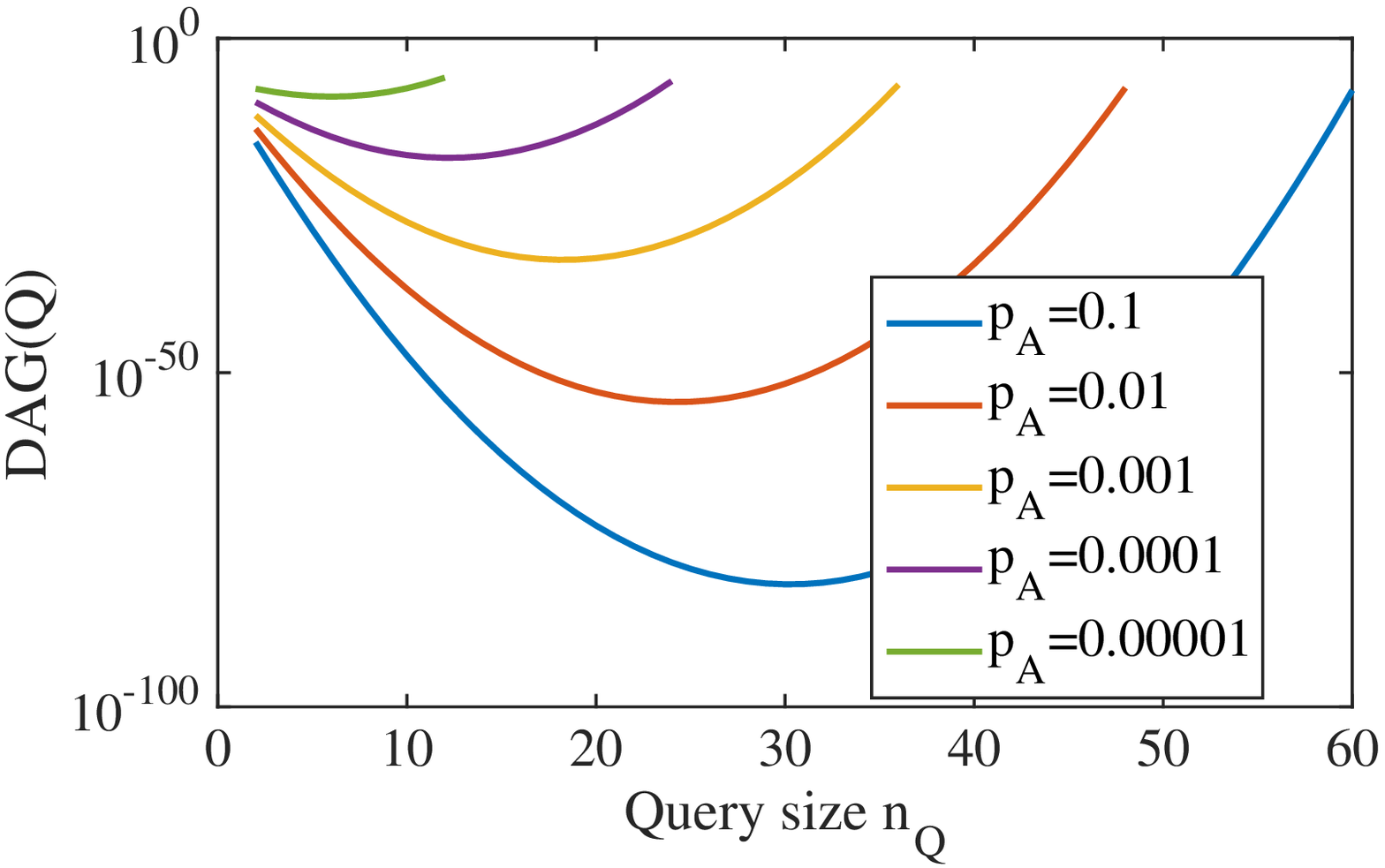}}
\subfigure[Varying $p_e$]{\label{edgesPr_partial_pe}\includegraphics[width=0.32\linewidth]{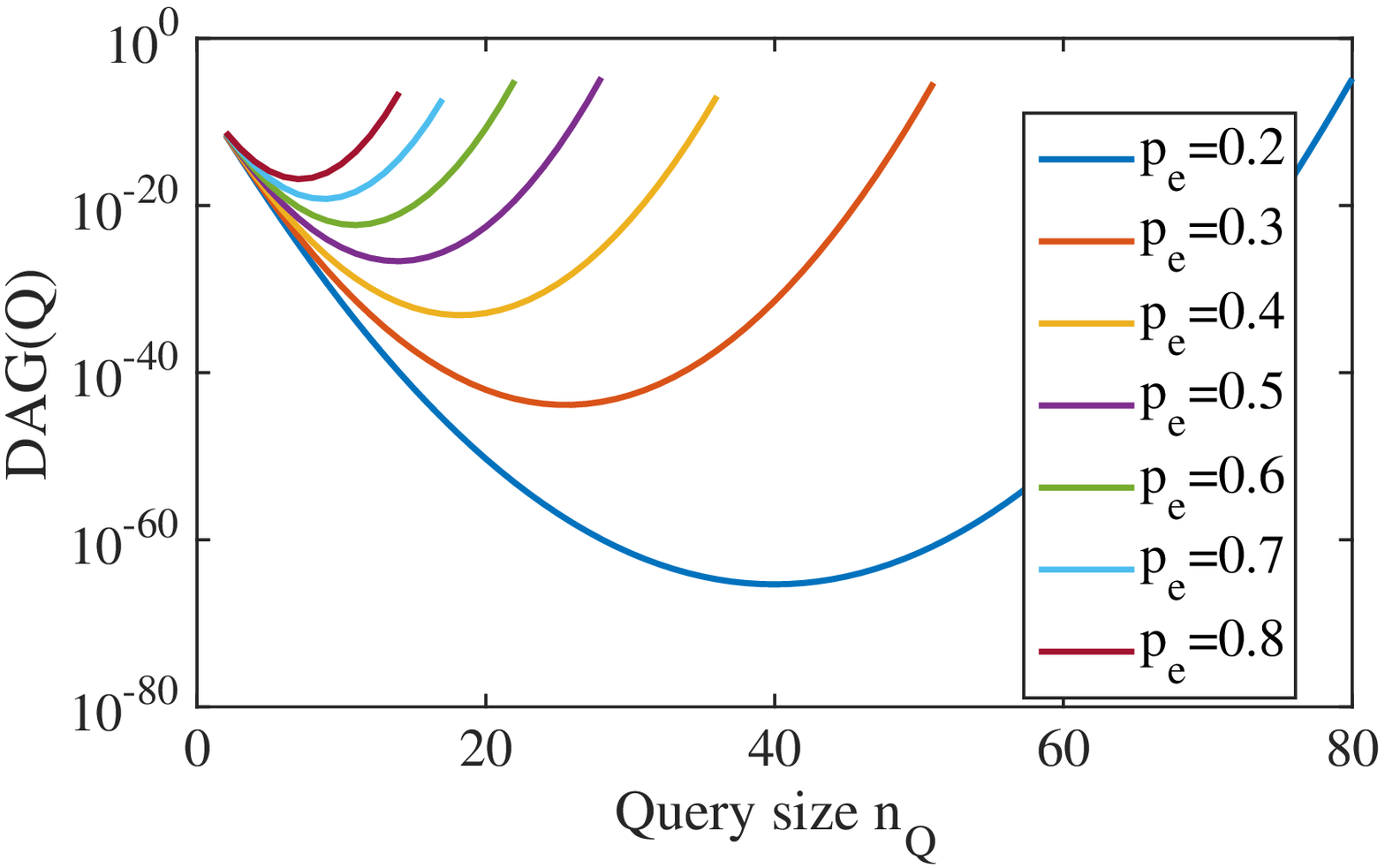}}
    \caption{The impact of $n_Q$ on \da gain for edges with probabilistic \& partial knowledge on \gnp.}\label{edgesPr_partial}
\end{figure*}
\begin{figure*}[tbhp]
\centering
\subfigure[$DAG(Q)$-$n_Q$, varying $p$]{\label{edgesPr_complete2_p}\includegraphics[width=0.32\linewidth]{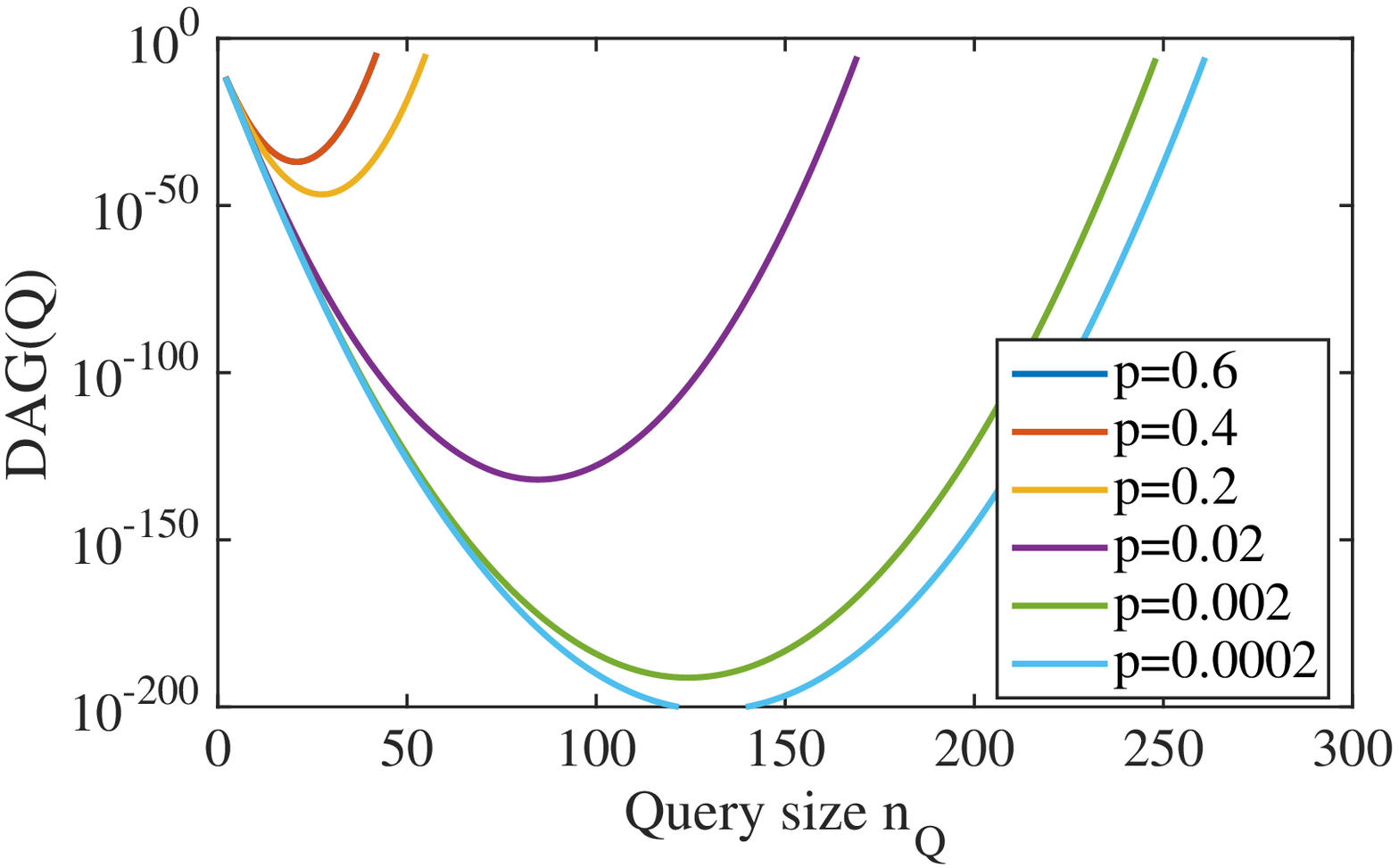}} 
\subfigure[$DAG(Q)$-$n_Q$, varying $p_A$]{\label{edgesPr_complete2_pA}\includegraphics[width=0.32\linewidth]{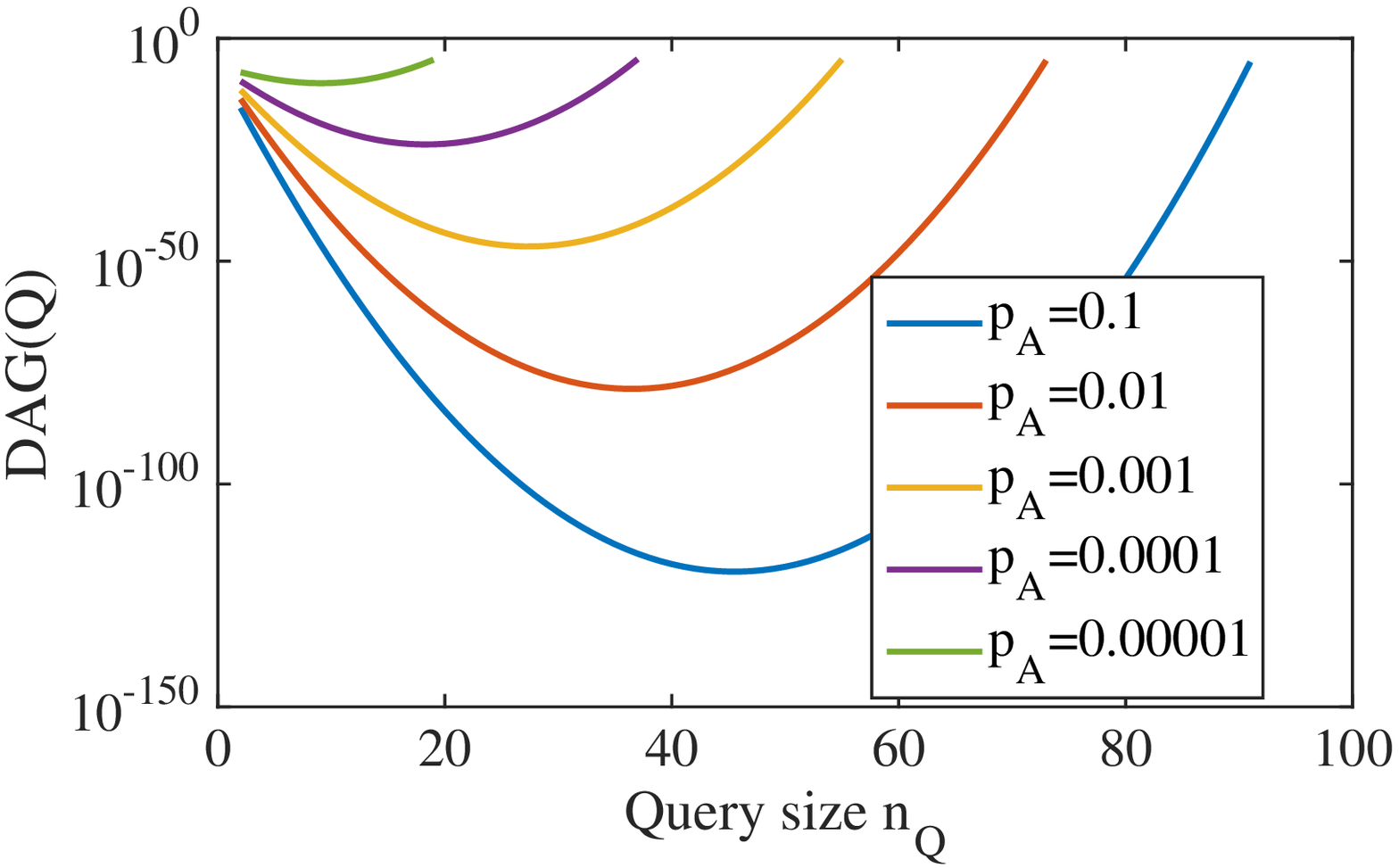}}
\subfigure[$DAG(Q)$-$n_Q$, varying $r$]{\label{edgesPr_complete2_r}\includegraphics[width=0.32\linewidth]{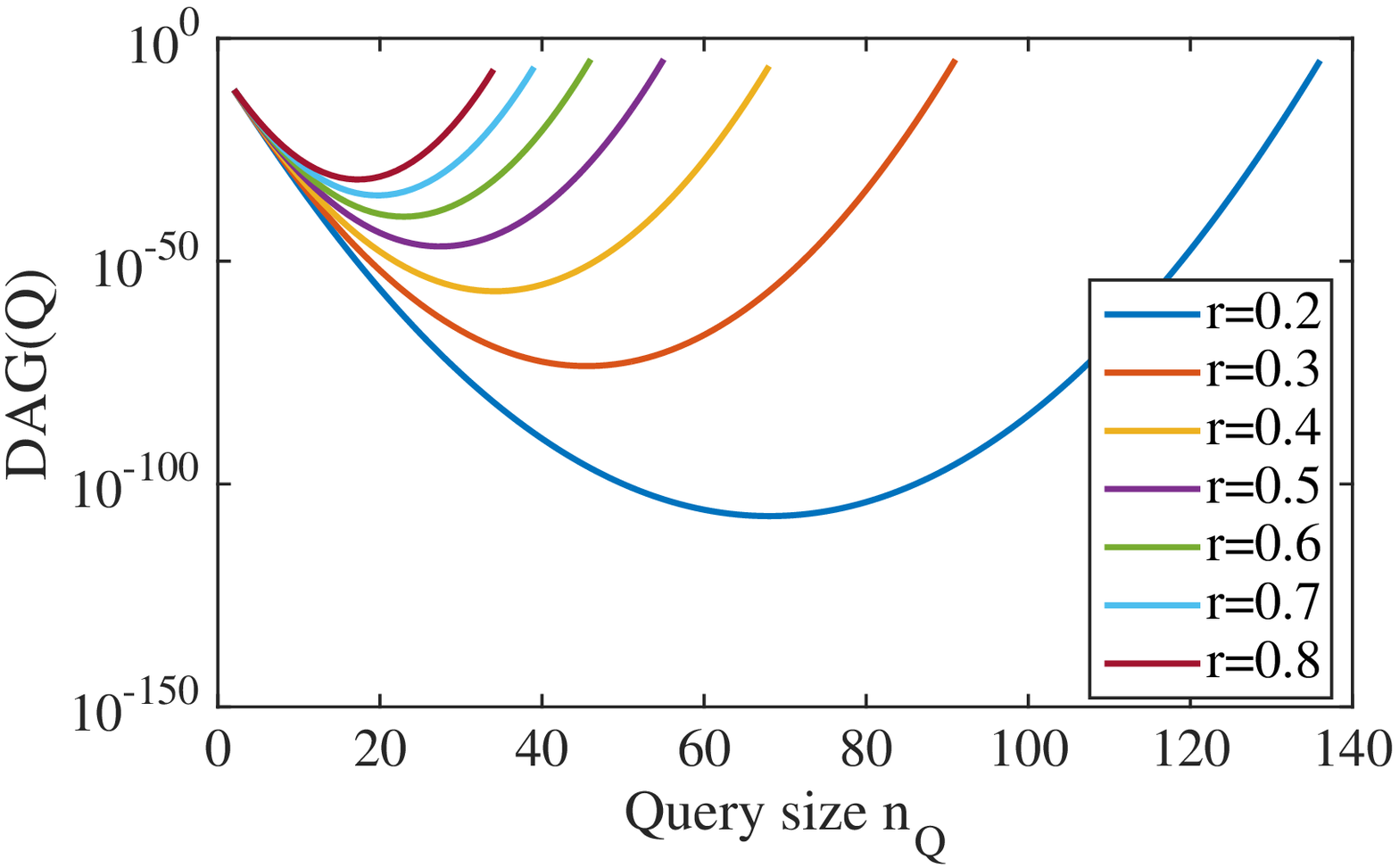}} \\
\subfigure[$DAG(Q)$-$r$, varying $p$]{\label{ePrCmp2_quality_p}\includegraphics[width=0.32\linewidth]{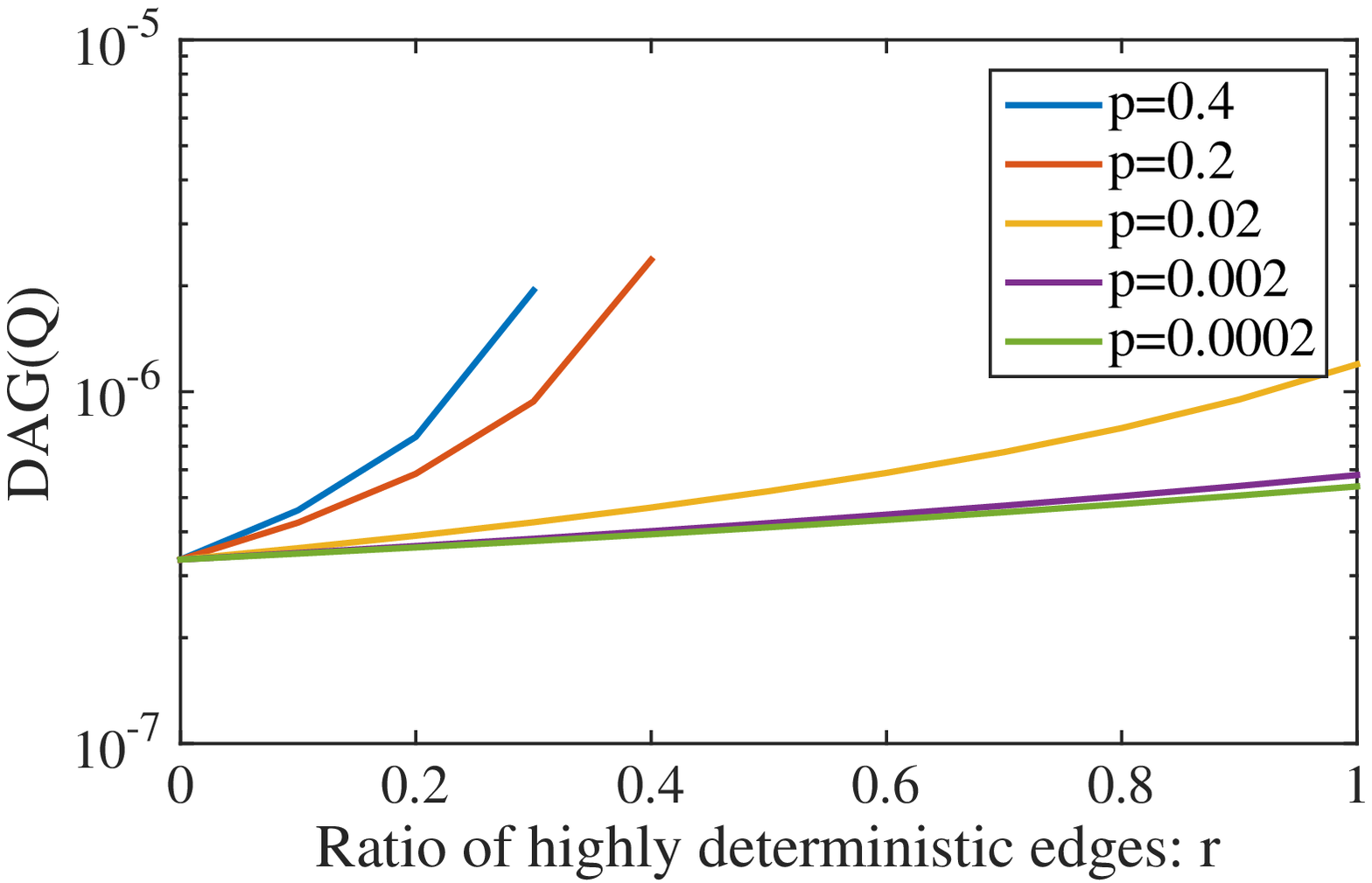}}
\subfigure[$DAG(Q)$-$r$, varying $n_q$]{\label{ePrCmp2_quality_nq}\includegraphics[width=0.32\linewidth]{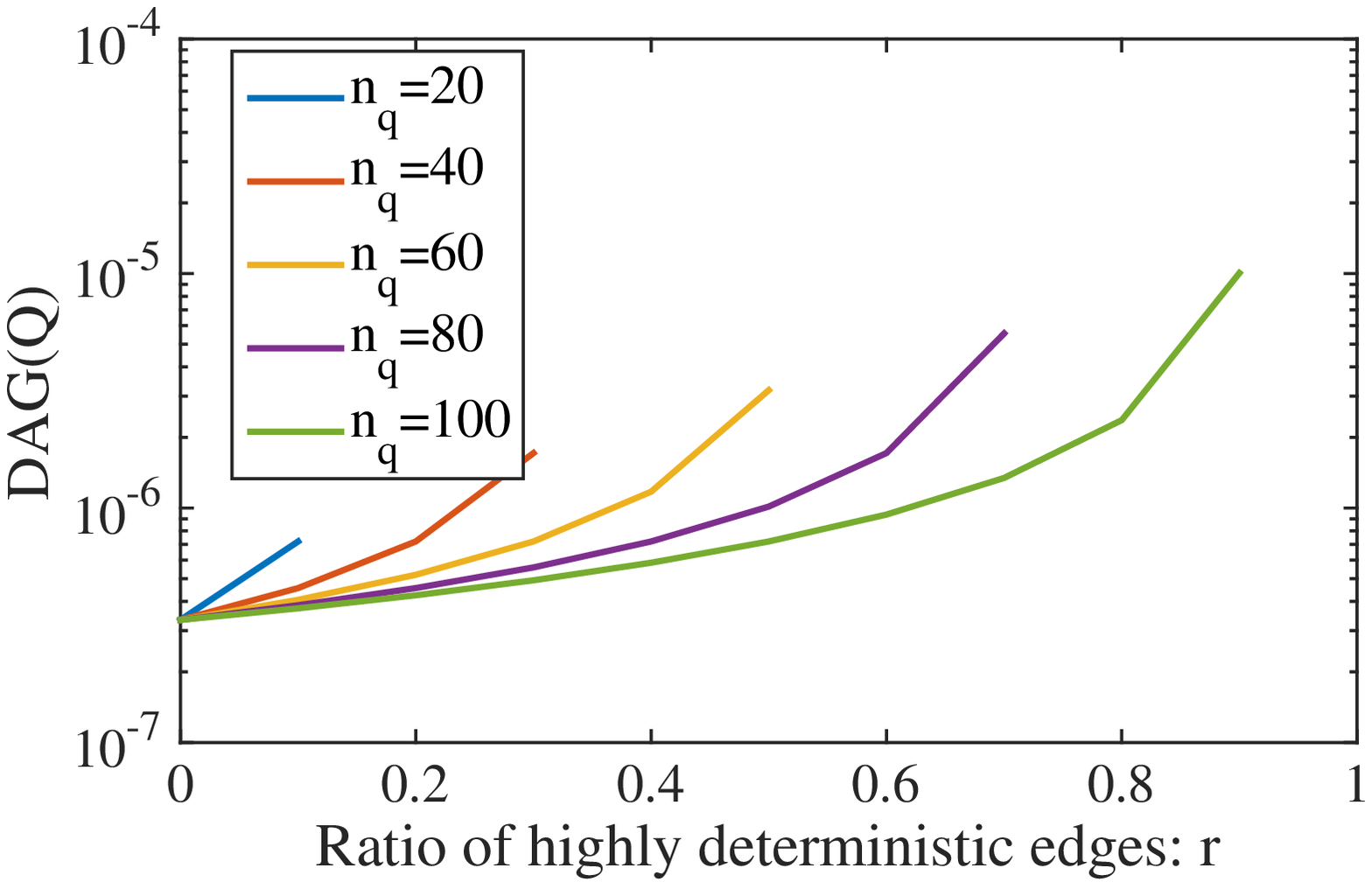}}
\subfigure[$DAG(Q)$-$r$, varying $p_A$]{\label{ePrCmp2_quality_pA}\includegraphics[width=0.32\linewidth]{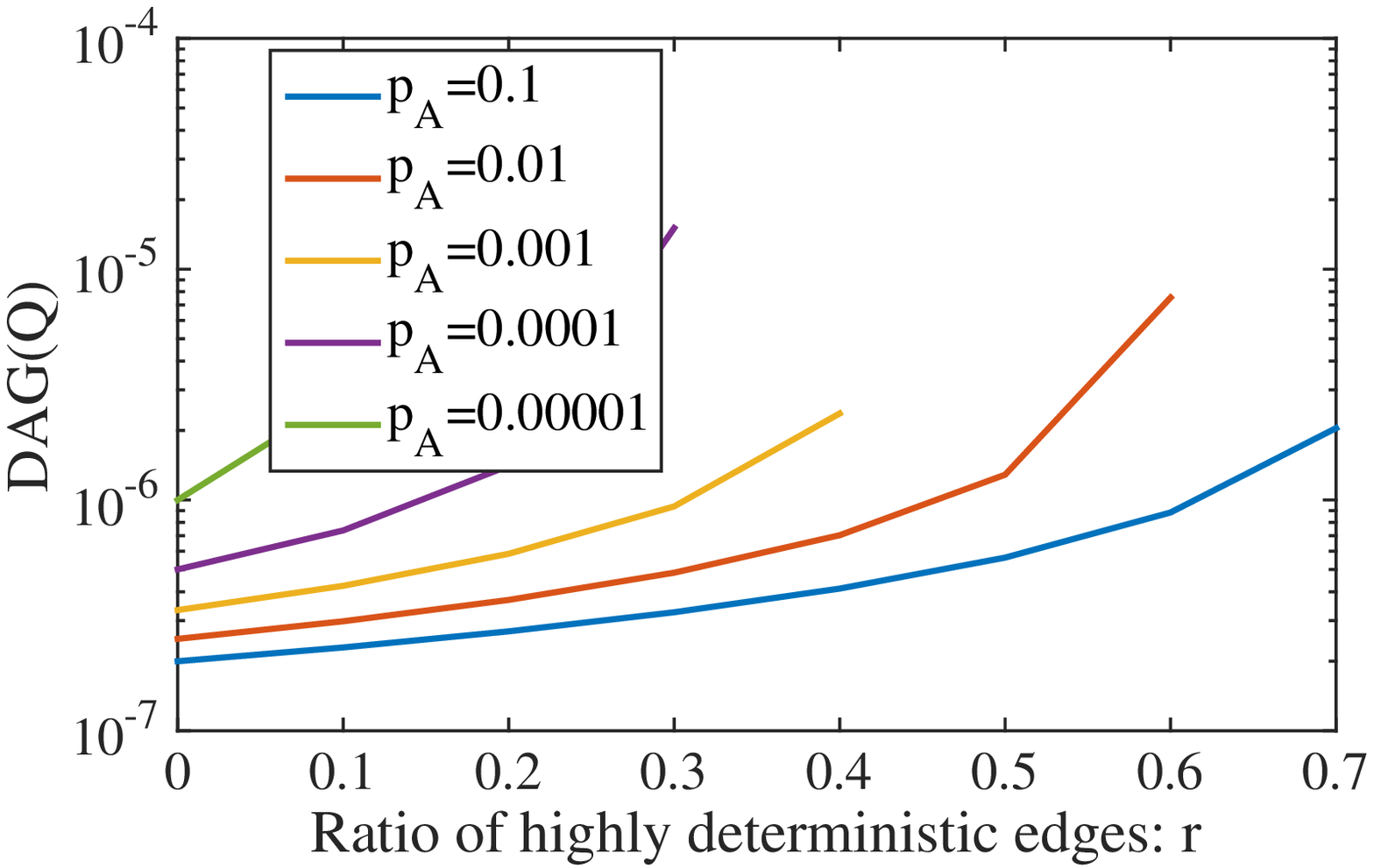}}
    \caption{The impact of $n_Q$ (Fig. a-c) and $r$ (Fig. d-f) on \da gain for probabilistic edges \& complete knowledge on \gnp.}\label{edgesPr_complete2}
\end{figure*}

As aforementioned we have listed the formulas to calculate $M_Q$ and thus $DAG(Q)$.
We study the impact of background knowledge on the \da gain
and plot it in Fig.~\ref{exact_partial}-\ref{edgesPr_complete2} for every case discussed above by varying all kinds of parameters. 
Please see Table~\ref{tab:bk} for a complete list.
The parameters used in these plots are listed in Table~\ref{tab:matlab-params}, and
we set a default value for each of them in order to control variables when plotting each of their influence on the results. 
We have tried all possible parameter values other than the default, and the results are similar under most settings.

When the \bk is quantified by the number of nodes $n_Q$,  in some settings
more \bk indeed results in more \da gain for the attacker, which conforms to our intuition. 
For instance, as shown in Fig. \ref{noisy_partial_pq}, \ref{noisy_complete_pq}), when the knowledge is noisy \& partial (or complete) and $p_q=0.5\sim 0.8$, $DAG(Q)$ is monotone increasing with respect to $n_Q$.
When the \bk is quantified by the ratio of highly deterministic edges $r$, $DAG(Q)$ is also monotone increasing with respect to $r$,
as indicated by Fig. \ref{edgesPr_complete2}(d-f). We see that in some cases, more adversarial knowledge implies higher risk of user identities being compromised.

However, this is not necessarily the case. Some of the curves
show a counter-intuitive relation between $DAG(Q)$ and $n_Q$, \eg most curves in Fig. \ref{exact_partial}. They have a \textit{transition phenomenon}:
with growing $n_Q$, $DAG(Q)$ first decreases until reaching a valley and then increases to the highest.
There are two \textbf{critical points}, valley point and vanish point, referring to the value of $n_Q$ where $DAG(Q)$ reaches the valley and the highest point respectively.
When $n_Q$ is greater than the vanish point, we get the estimated number of matches $M_Q<1$ according to the formulas. 
Such a result does not make sense so we do not plot it on the curves. 
In real-world attacks, when the \bk reaches the vanish point, the attacker finds the only one match and thus users are de-anonymized successfully.
Furthermore, the curves reveal that the position of the critical points are decided by multiple parameters including $p,p_q, p_A,r$.

The occurrence of \tp can be explain by Theorem \ref{theorem}. The match number $M_Q$ which determines \da gain is influenced by
two factors, the size of mapping space $\binom{n}{n_Q}n_Q!$ and the matching \pr $Pr(Q\simeq I)$.
The mapping space increases with growing $n_Q$ owing to an increasing number of node permutations.
The matching \pr decreases with respect to $n_Q$ as a randomly selected candidate needs to 
meet more and more requirements to be a match of $Q$. The two factors have the opposite effects on $M_Q$ so there is a possibility that
a \tp occurs. 

We also plotted (though not presented here) the relation between \da gain and $p_q$ with fixed $n_Q$. 
The results conforms to the intuition, \ie a denser $Q$ would contribute to a more successful \da attack.
Therefore, we reach the conclusion that more \kg does imply more \da gain in some settings, but that might not always be the case.

\begin{table}[!t]\small
\centering
\begin{tabular}{l|c|c|c|c}\hline
Fig. & Correctness & Completeness & Attributes & Confidence \\\hline\hline
\ref{exact_partial}(a-c) & exact & partial & included & deterministic \\\hline 
\ref{exact_complete}(d-f) & exact & complete & included & deterministic \\\hline
\ref{noisy_partial}(a-d) & noisy & partial & included & deterministic \\\hline 
\ref{noisy_complete}(e-h) & noisy & complete & included & deterministic \\\hline 
\ref{edgesPr_partial} & N/A & partial & included & probabilistic \\\hline 
\ref{edgesPr_complete2} & N/A & complete & included & probabilistic \\\hline 
\end{tabular}
\caption{Figures and background knowledge types for \gnp graphs.}
\label{tab:bk}
\end{table}





\section{Theoretical Analysis on Power-Law Graphs}
\label{sec:power}
In this section, we study the relation of \da gain and background knowledge when $G$ conforms to the power-law model.
A power-law graph can be generated with a given degree sequence that
has a power-law distribution \cite{chung2002connected}. Specifically,
$n(d)=\alpha d^{-\beta}$, where $n(d)$ represents the number of nodes with degree $d\in\mathbb{N}_+$, 
$\alpha,\beta$ are predefined parameters, and $\beta>2$.
It is tricky to give a formula of $M_Q$, but we derive a lower bound for it 
for the case of exact and partial knowledge with attribute ignored.
\begin{theorem}
$M_Q\geq \binom{n}{n_Q}n_Q!(\frac{\beta-1}{n(\beta-2)})^{m_Q}$.
\end{theorem}
The proof is given in Appendix \ref{app2}. We can use this result to estimate the upper bound of $DAG(Q)$.

\begin{figure}[tbhp]
\centering
\subfigure[Varying $\beta$]{\label{pl_upper_beta}\includegraphics[width=0.32\textwidth]{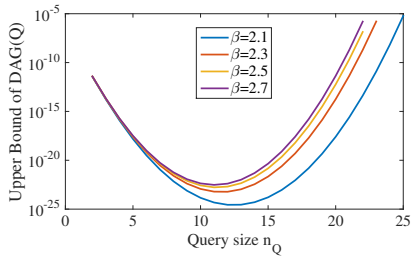}} 
\subfigure[Varying $p_q$]{\label{pl_upper_pq}\includegraphics[width=0.32\textwidth]{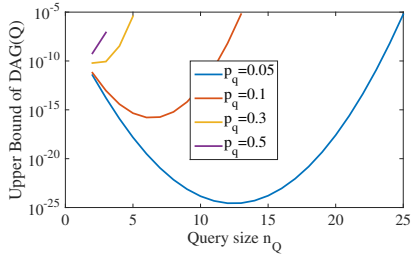}}
    \caption{The impact of $n_Q$ on the privacy lower bound for the exact \& partial knowledge on power-law graphs}\label{pl_upper}
\end{figure}

\textbf{Analytical Results:}
We set $p_q=0.05$ as default and set other parameters as given in Table \ref{tab:matlab-params}.
As shown in Fig. \ref{pl_upper}, in some settings there
is a transition phenomenon in the relation between the upper bound of $DAG(Q)$ and $n_Q$ for power-law graphs, 
but in other settings (\eg $p_q=0.3\sim0.5$) there is no \tp (Fig. \ref{pl_upper_pq}).

\section{Simulations}
\label{sec:experiment}
In this section, we present our simulations on both synthetic data and real network datasets.
We would explain the relation observed between \bk and \da gain based on the results obtained.

\subsection{Methodology}
Due to the hardness of \sm and subgraph counting, many previous works focus on designing approximation algorithms \eg \cite{slota2013fast,babai2015graph,chakaravarthy2016subgraph}.
The biggest challenge in the experiment is to estimate the match number $M_Q$, which relies on the NP-complete subgraph counting problem.
We implement a state-of-the-art approximate algorithm in \cite{chakaravarthy2016subgraph}, which still has \textit{exponential} time and space complexity.
Although it only applies to undirected graphs and query graphs with treewidth no more than 2,
it is so far the work that has the \textit{least} restricts about the size and structure of queries. 
On the contrary, other approximation algorithms such as \cite{slota2013fast} are usually limited to 
very small-sized query graphs (less than 10 nodes for example), or query graphs with a specific structure like triangle, circle or tree.

\begin{figure*}[tbhp]
\centering
\subfigure[\gnp graphs]{\label{starER}\includegraphics[width=0.4\linewidth]{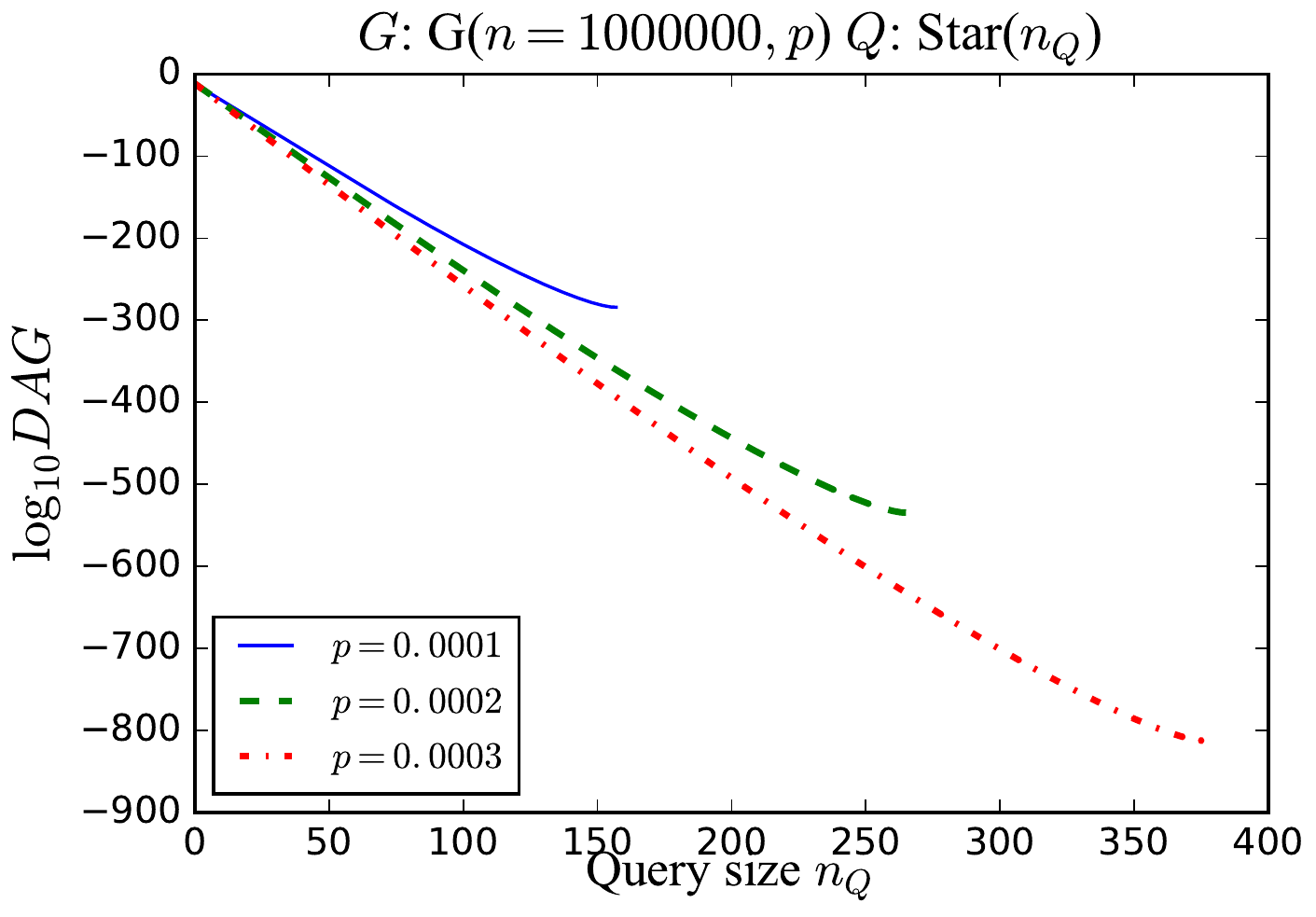}} 
\subfigure[Power-law graphs]{\label{starPL}\includegraphics[width=0.4\linewidth]{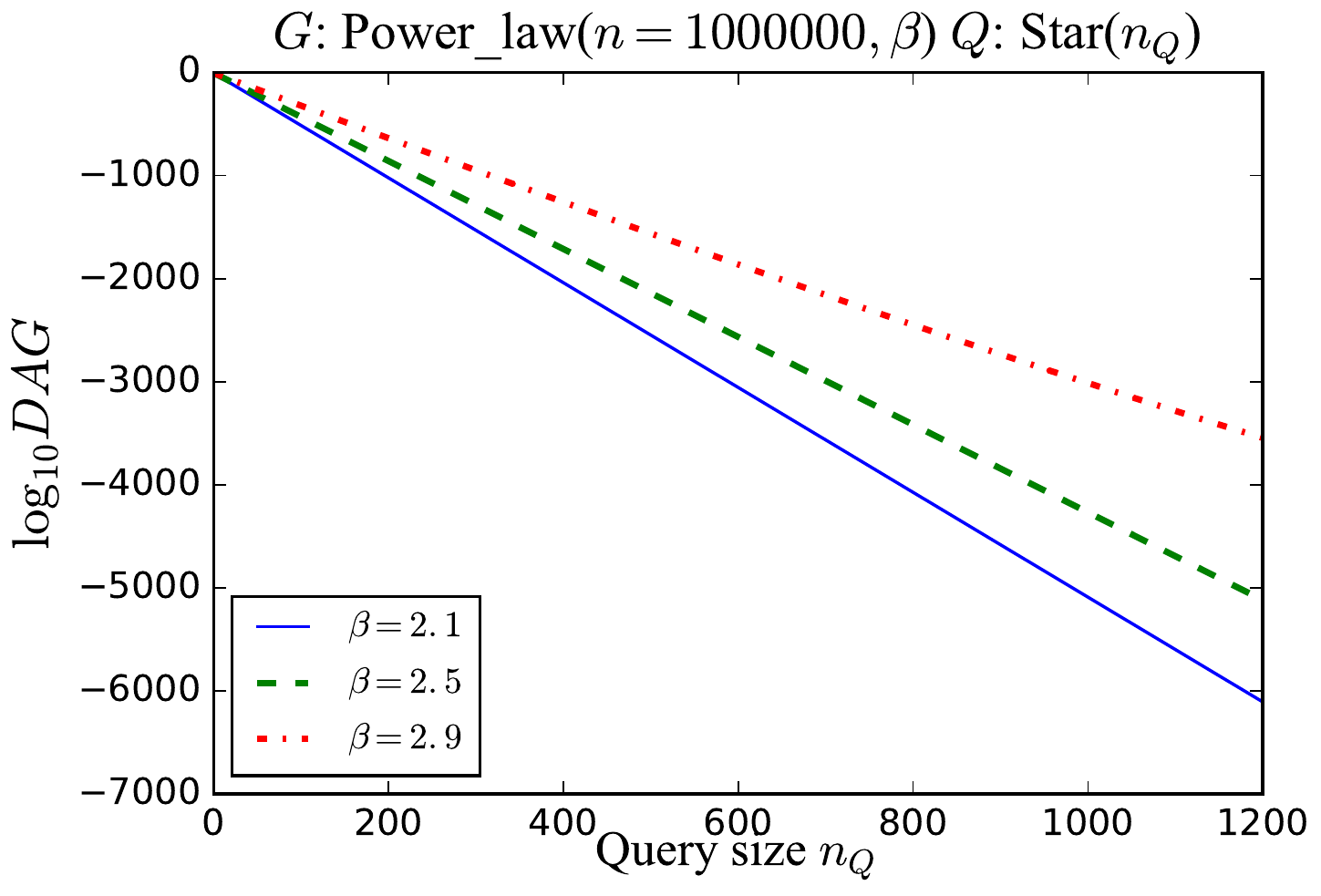}}\\ 
\subfigure[Real networks]{\label{starReal}\includegraphics[width=0.4\linewidth]{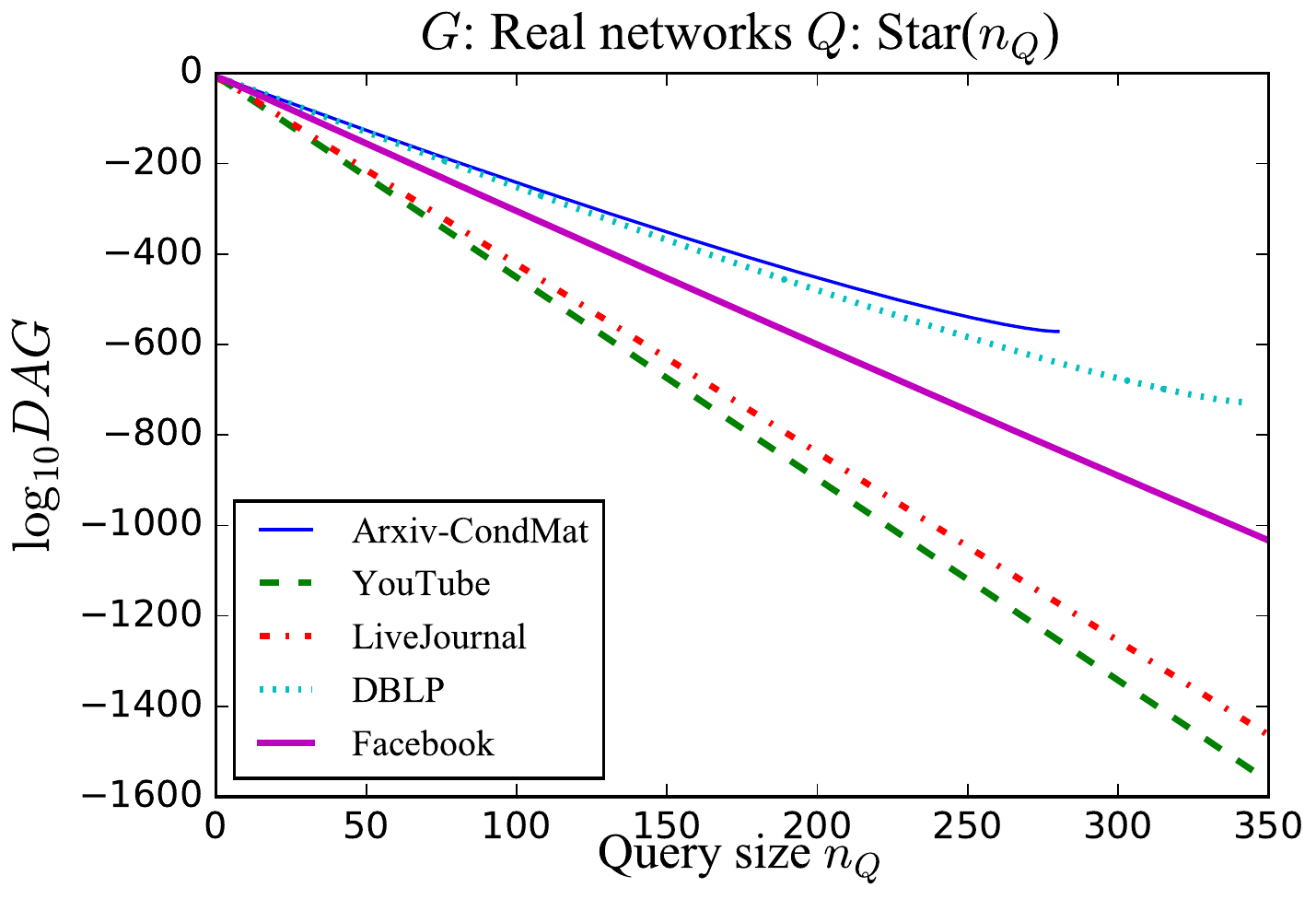}} 
\subfigure[Degree distribution]{\label{ddistri}\includegraphics[width=0.42\linewidth]{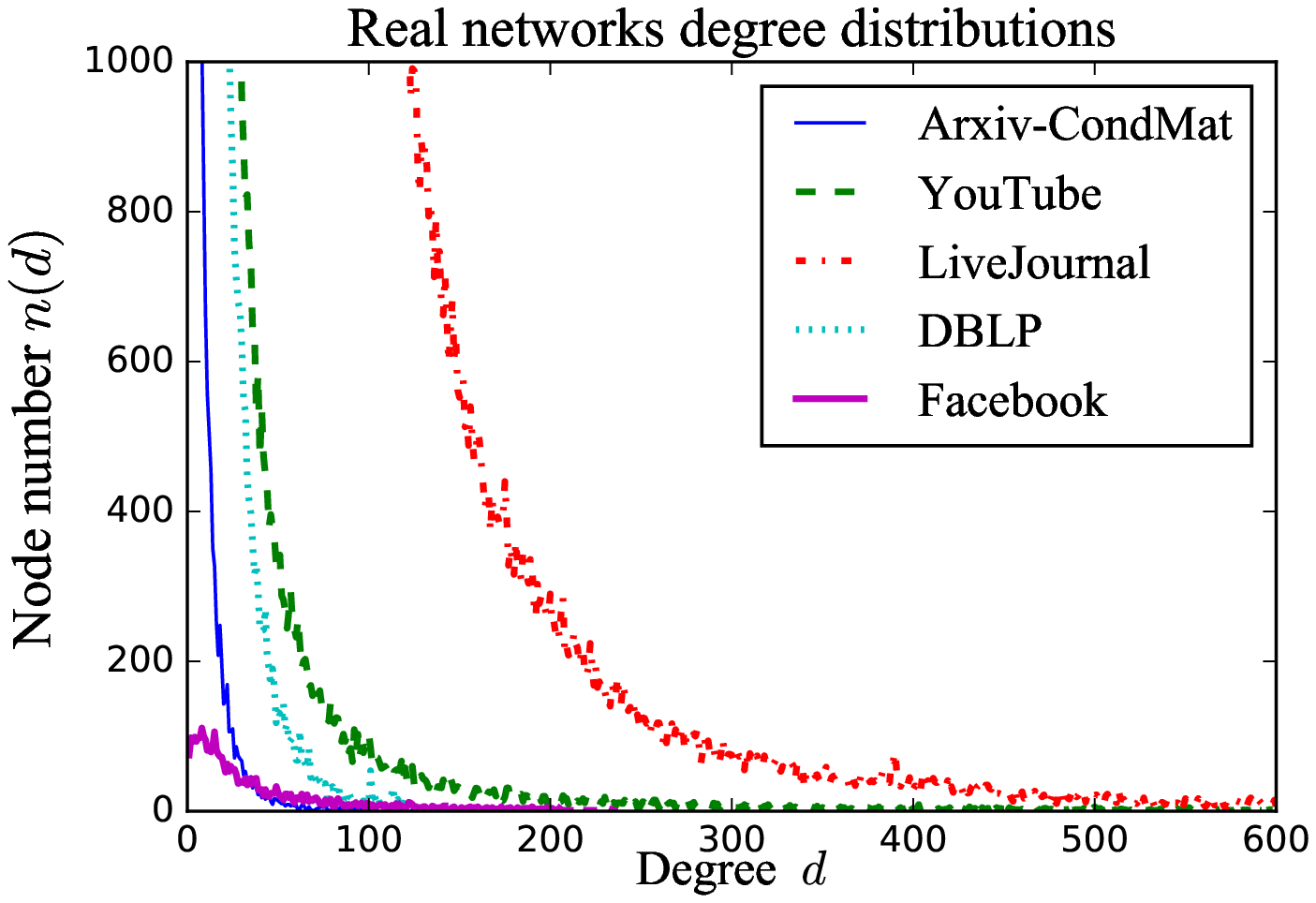}} 
    \caption{\textbf{The impact of $n_Q$ on $DAG(Q)$ for star queries on synthetic/real networks (Fig. a-c), and degree distribution of real networks (Fig. d)}: The greater $n_Q$ is, the smaller $DAG(Q)$ is.}
    \label{star-query}
\end{figure*}

\textbf{Datasets:} We verify our claims on both synthetic data and real network data.
For synthetic data, we generate random power-law graphs with $n=100K$ and different $\beta$.
Experiment of querying on \gnp graphs is not presented here for space limit 
(it has been analyzed and plotted in Section \ref{ER-analysis}).
For real datasets, we use both online social networks (Facebook, YouTube, LiveJournal) and collaboration networks (DBLP, Arxiv Cond-Mat)
(see Table \ref{tab:real-data} for statistics). They have power-law like degree distribution as shown in Fig \ref{ddistri}.

The experiment on each dataset consists of \textbf{4 steps}:
1) preprocess the original graph $G$, 
2) generate query graphs $Q$ by varying $n_Q$ and $m_Q$ respectively,
3) query $G$ with $Q$, calculate $M_Q$ and $DAG(Q)$, and
4) plot the relation between $DAG(Q)$ and $n_Q$ or graph density $p_q$ ($=2m_Q/n_Q(n_Q-1)$).

In the second step above, query graphs are generated by randomly extracting ego networks from $G$. 
De-anonymizing ego networks is also commonly researched in the literature \cite{zhou2008preserving,wang2013outsourcing}.
We distinguish ego networks as \textit{star} and non-stars as they have different influence on the matching results.
Star queries usually have large quantities of matches in $G$ due to their considerable automorphisms.

\begin{figure*}[tbhp]
\centering
\subfigure[$DAG(Q)$-$n_Q$ (synthetic)]{\label{pri_nQ}\includegraphics[width=0.4\linewidth]{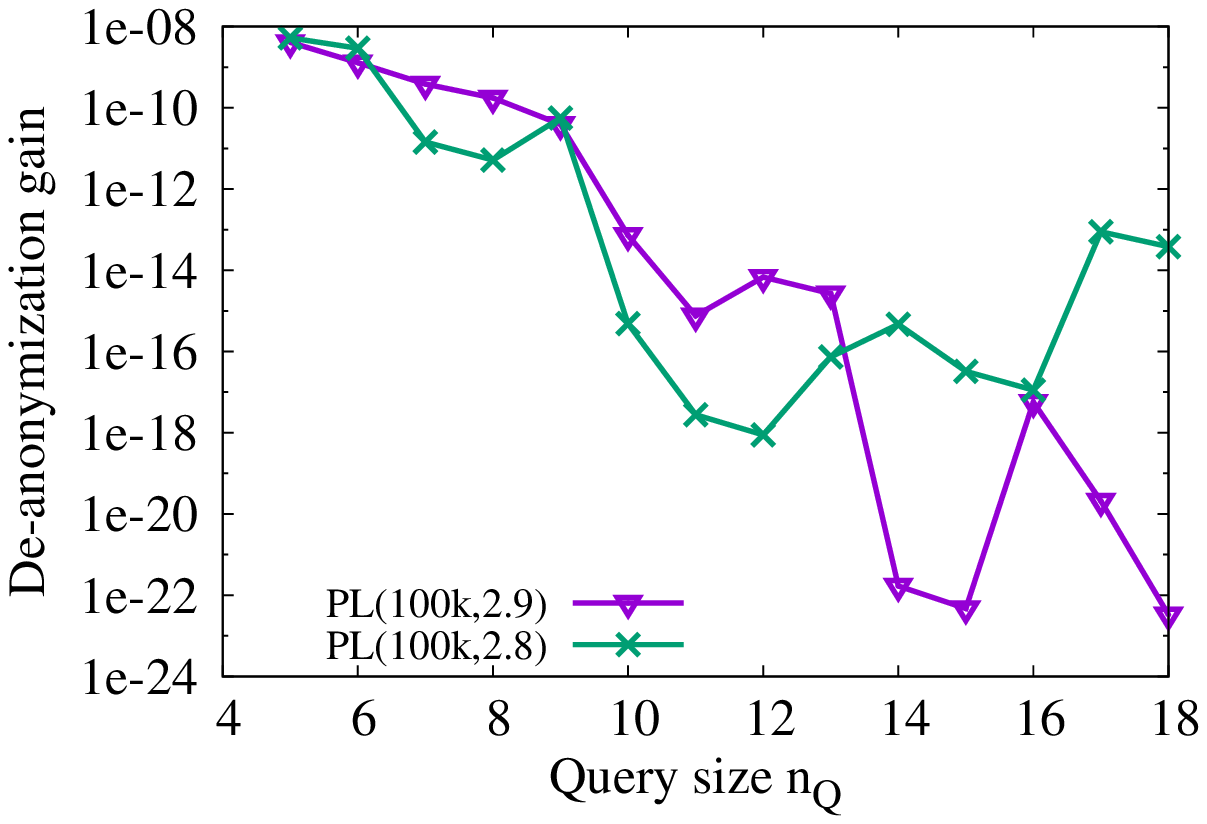}} 
\subfigure[$DAG(Q)$-$n_Q$ (real)]{\label{pri_nQ_real}\includegraphics[width=0.4\linewidth]{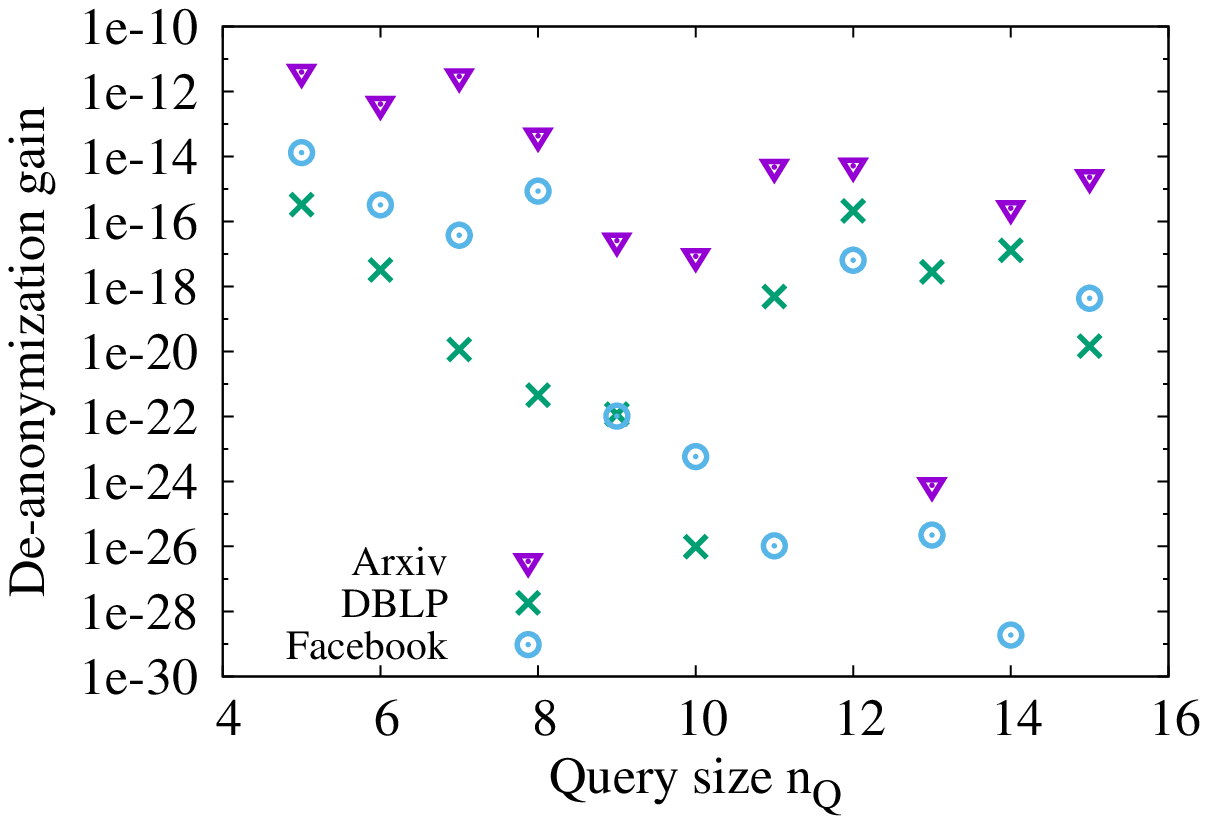}}\\
\subfigure[$DAG(Q)$-$p_q$ ($n_Q=5$)]{\label{pri_pq5}\includegraphics[width=0.4\linewidth]{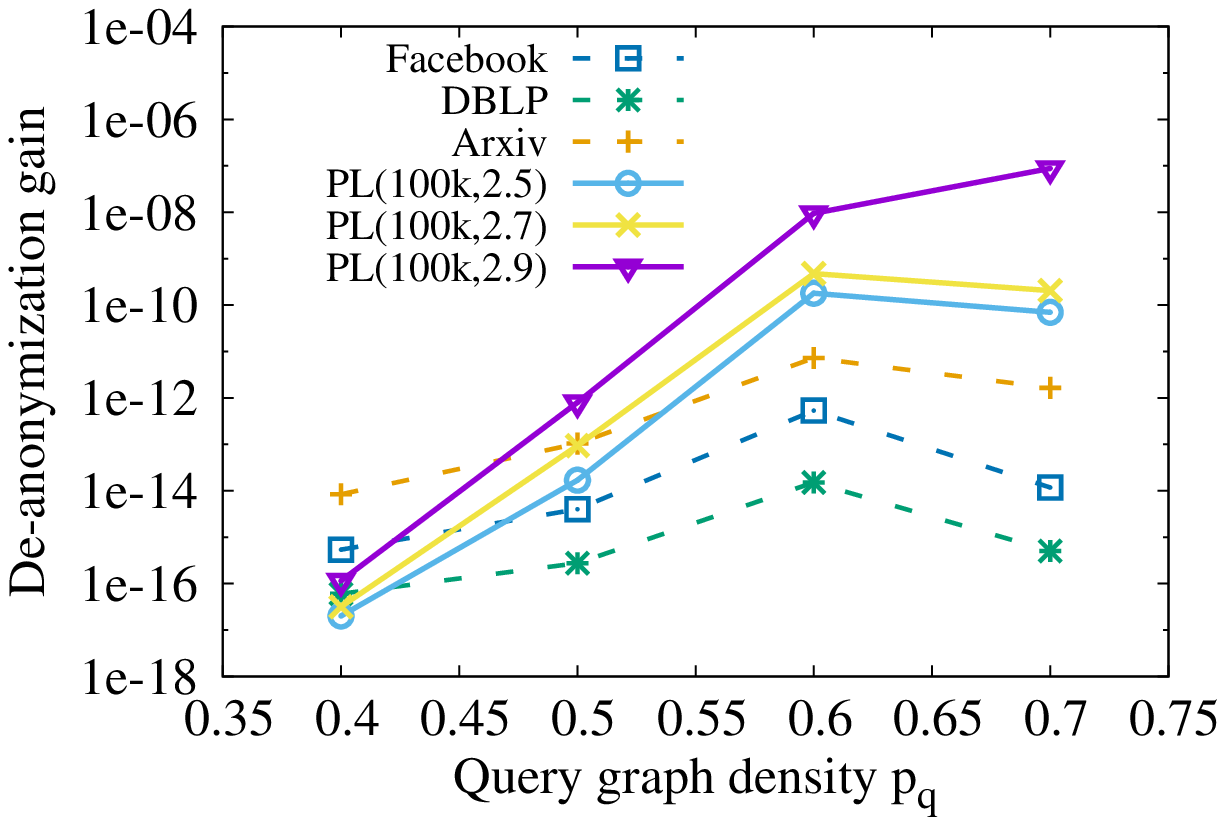}}
\subfigure[$DAG(Q)$-$p_q$ ($n_Q=7$)]{\label{pri_pq7}\includegraphics[width=0.4\linewidth]{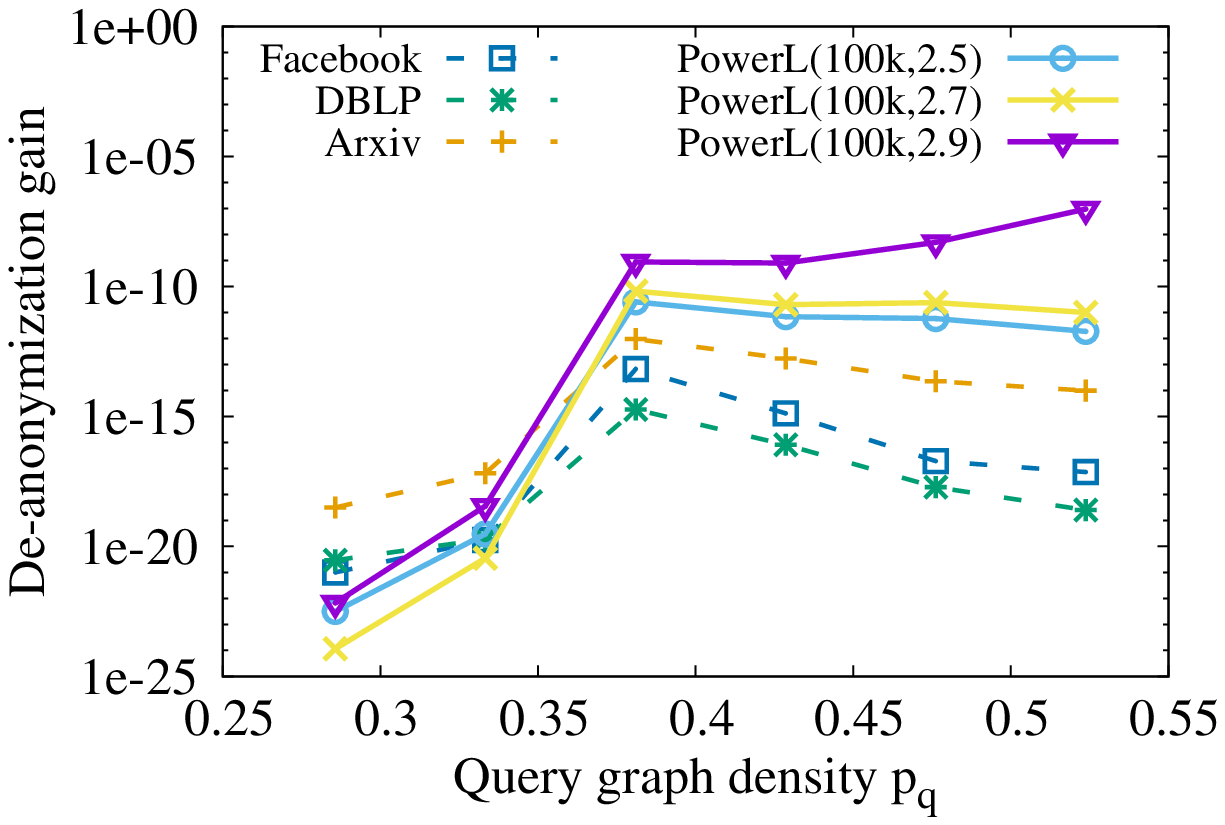}}
    \caption{\textbf{The impact of $n_Q$ and $p_q$ on $DAG(Q)$ for non-star queries on synthetic and real networks:} $DAG(Q)$ decreases with growing $n_Q$. When $p_q$ increases, $DAG(Q)$ first rises and then falls off a bit.}
    \label{fig:pl}
\end{figure*}

\begin{table}\small
\centering
 \begin{tabular}{ c | c | c | c | c | c}
    \hline
     & DBLP & Arxiv & Facebook & YouTube & LiveJournal  \\ \hline \hline
    $n$ & 0.3M & 23K & 4K & 1.1M & 4M \\ \hline 
    $m$ & 1M & 93K & 88K  & 3M & 34.7M  \\ \hline 
 \end{tabular}
 \caption{Real collaboration/online social network datasets}
\label{tab:real-data}
\end{table}

\subsection{Experimental Results and Analysis}

The majority of ego networks we sampled are stars (especially for power-law graphs, the ratio is $60\sim 90\%$). The results for star queries
are displayed in Fig. \ref{star-query}(a-c), which reveals that de-anonymizing $Q$ of greater size is more difficult. This is because stars of larger size have more automorphisms which results in more matches found in $G$. The curves in the figure end when $n_Q$ exceeds the highest degree.
For non-star queries, the results are presented as follows.

\textit{Knowledge quantified by $n_Q$ (quantity):}
The relation between \da gain and $n_Q$ for synthetic data is plotted in Fig.~\ref{pri_nQ}. 
It reveals that $DAG(Q)$ first descends and then fluctuates or picks up a bit when $n_Q$ grows.
We did not obtain the experiment data for $n_Q>18$ due to its excessive demand for memory. Though we have not clearly observe the \tp yet, we can at least learn from this figure 
that more \bk does not always induce more \da gain in some cases.
Similar phenomena are also found for real networks, as depicted by Fig. ~\ref{pri_nQ_real}. The curves are not smooth due to two possible reasons:
1) real networks do not exactly follow the power-law model and 
2) the approximate subgraph counting algorithm we utilize does not output exact match numbers.

\textit{Knowledge quantified by $p_q$ (quality):}
Fig. \ref{pri_pq5} and \ref{pri_pq7} display the relation between $DAG(Q)$ and $p_q$. 
It is shown that sometimes $DAG(Q)$ first increases then slowly decreases again with growing $p_q$.
At first, the denser $Q$ is, the more ``special'' it is, so fewer matches are found and it is easier to pinpoint each user.
However, when $p_q$ reaches a threshold, $Q$ has more automorphisms when $p_q$ is greater, 
\eg, $Q$ has the most number of automorphisms when it is a clique. This is a possible explanation of the fall after rise of $DAG(Q)$.
Therefore, we can learn that more \ak does not always bring more \da gain.
Some may question why few data points are plotted in the figures. 
This is because $p_q$ is determined by $m_Q$, and the algorithm we use does not support querying dense graphs.

\subsection{Implications}
\label{subsec:implic}

\textbf{Medium-density queries are more powerful:} Graph automorphism makes it harder to tell individual nodes apart and exactly de-anonymize them, even when the attacker finds the right subgraph in $G$ corresponding to $Q$. An ego network has the most number of automorphisms when
it is either very sparse (close to a star) or very dense (close to a clique).  As revealed in Fig. \ref{pri_pq5} and \ref{pri_pq7}, 
the \da gain first increases to the peak and then gradually drops again. 
Namely, de-anonymizing adversarial graphs of medium density might be the easiest. 
By the cask effect, the data protector can design perturbation methods to lower the peak so as to improve the overall data privacy.

\textbf{Attack-resistance of data:} No matter if there is a \tp, the vanish point reflects the attack-resistance property of the published graph $G$, since it is the point with the worst user identity leak (the target users are all successfully de-anonymized). A greater vanish point implies that $G$ is more resistant to de-anonymization attacks, that is, the attacker needs to collect more \bk to recover users' identities.
The position of the vanish point is up to the properties of $G$ itself and the parameter assumptions about $Q$ (recall Section \ref{ER-analysis}).
An implication for data publishers is that
changing $G$'s properties during data perturbation to defer the vanish point may also help reduce the risk of privacy leak.

\textbf{Break knowledge and attack better:} The difficulty of de-anonymization is two-fold: the computational complexity of subgraph matching and the indistinguishability of the matches. In the settings with a \tp, the attacker might achieve better attacker results, \ie pinpoint users with a higher \pr, if she breaks down her \bk graph into several smaller graphs and finds their matches separately. 
For example, when $n_Q$ is around the peak point, this strategy might make the \da more efficient and accurate.
However, the attacker should not reduce her knowledge when $n_Q$ is near the vanish point, or she would achieve worse attack result.
Though the effectiveness of the strategy still needs verifying, our work might help the attacker to make better use of the \bk for \da attack. 
The question when and how to break the knowledge is saved for our future work.

\section{Discussion}
\label{sec:discuss}



\textbf{More complex \da gain definitions:} 
Sometimes user identity privacy is disclosed when 
the attacker knows that a user belongs to a community (such as a drug rehabilitation or disease treatment group),
even though the user is not exactly de-anonymized at the individual level. We refer to it as community level \da.
Since the mapping between $Q$ and $I$ does not matter, the number of \textit{matched communities} $C_Q$ 
(defined as follows) is more relevant than $M_Q$ to privacy measure.
\begin{definition}[Matched Community]
Given $G$ and $Q$, a matched community is a subset $V_s\subseteq V$ \st 
there is a bijective mapping $f:V_Q\rightarrow V_s$ that satisfies
$$\forall e(u,v)\in E_Q, e(f(u),f(v))\in E.$$
\end{definition}
There is no exact relation between $C_Q$ and $M_Q$, but  $C_Q$ can be counted by checking the nodes of every match in $\mathcal{M}_Q$,
which relies on solving the NP-complete \sm enumeration problem. 

\textbf{$\ell$-Indistinguishability:} 
We can define $\ell$-Indistinguishability to measure the privacy of the published graph $G$, which is similar to  $k$-anonymity \cite{sweeney2002k}.
Suppose $\ell=M_Q$ matches are found by querying $G$ with $Q$,
we say $G$ satisfies \textit{weak} $\ell$-indistinguishability under the attack of $Q$.
However, there might be overlapped nodes in the $\ell$ matches \cite{cheng2010k}.
To overcome this weakness, 
we can construct a new query graph $Q'$ which consists of $\ell$ disjoint copies of $Q$, and use $Q'$ to query $G$.
If $M_Q'\geq 1$, then we say \textit{strong} $\ell$-indistinguishability is achieved.


\section{Related Works}
\label{sec:related}
\subsection{Social Network Anonymization and De-Anonymization}
The fight between de-anonymization and anonymization is becoming more and more fierce.
Some researches aim at attacking/protecting an individual's privacy on the basis of $k$-anonymity \cite{sweeney2002k}.
For example, $k$-neighborhood anonymity
\cite{zhou2008preserving,wang2013outsourcing} was proposed to protect against 1-neighborhood attack and $1^*$-neighborhood attack, 
and $k^2$-degree anonymity \cite{tai2011privacy} is proposed to defend against friendship attack, where the attacker is assume to know the degrees of two users who are friends.
Some methods add great perturbations to the published graph to achieve
$k$-candidate anonymity \cite{hay2007anonymizing}, $k$-automorphism \cite{zou2009k}, or $k$-isomorphism \cite{cheng2010k}, 
yet cause the loss of data utility/quality.

Other related works focus on graph mapping attacks, also referred to as structure-based de-anonymization,
in which the attacker de-anonymize users  by mapping one network he possesses to the published network.
Most of these attacks are seed-based, including \cite{backstrom2007wherefore,narayanan2009anonymizing, 
ji2015your}. Here seed users refer to outstanding users, such as users with very high degrees like celebrities.
There are also works that do not need seed users, \eg \cite{pedarsani2013bayesian,ji2014structural,Qian2016},
which are based on Bayesian model, optimization and  knowledge graph model respectively.
As shown by \cite{ji2014structural}, most existing social network datasets are de-anonymizable partially or completely, 
and there is no effective countermeasure proposed yet.

\subsection{Subgraph Matching/Isomorphism}

As aforementioned, the subgraph matching problem is NP-complete \cite{cook1971complexity}. However, it has a wide variety of applications including biological networks \cite{he2008graphs},   
 knowledge bases \cite{kasneci2008naga}, 
 and program analysis \cite{zhang2010sapper}.  

A widely accepted approximation in subgraph matching is to perform prune-and-search by indexing the graph data. 
Such approaches can be classified based on how indexing is performed: edge index \cite{prud2008sparql}, 
frequent subgraph index \cite{giugno2002graphgrep},  
and neighborhood index
 \cite{zou2009distance}.  
To our knowledge, the STwig deployed on Trinity Memory Cloud  \cite{sun2012efficient} are the most efficient; the run time is several seconds when a graph with  tens of nodes and edges is queried in a graph of size 1 billion.

Extended from subgraph matching, the subgraph counting problem is more related to our paper but even harder to solve.
Most of existing algorithms can only estimate
the count when the subgraph has a small size or a particular structure like tree, cycle or triangle \cite{slota2013fast,chakaravarthy2016subgraph}.
Slota \etal \cite{slota2013fast} applied the color coding technique \cite{alon2008biomolecular} to approximately count non-induced occurrences of tree subgraphs, yet the algorithm can only count for subgraphs with at most 12 nodes.
A state of the art approximate subgraph counting algorithm was proposed by \cite{chakaravarthy2016subgraph} in 2016,
which can efficiently find matches for query graphs with treewidth no more than 2.

\section{Conclusion}
\label{sec:conclusion}

This work presents a comprehensive analysis on the impact of the attacker's \bk on \da gain for \sn \da.
First of all, we elaborately categorize \bk in multiple dimensions in terms of its properties, and analyze how the type of \bk influences the definition of ``matching'' and thus the result of \da.
We quantify background knowledge by both quantity and quality and introduce a definition for \da gain.
Then we present a detailed theoretical analysis on the relation between \bk and \da gain for network data of two popular data models (\gnp and power-law),
which reveals that in some settings \da gain is not necessarily monotone increasing with the amount of \bk.  
Despite the hardness of subgraph counting, we conduct simulations on both synthetic and real networks, which further verifies our claim and leaves implications for the data protector and the attacker.

\appendices
\section{Proof of Theorem 1}
\label{app1}
\begin{IEEEproof}
The idea behind this formula is similar to exhaustive search. As shown in Fig. \ref{subg-match}, 
the number of possible candidates like $I$ is $\binom{n}{n_Q}n_Q!$. Since $G$ is a random graph, 
we can compute the probability of $Q$ and $I$ being a match (defined later) and then use it to estimate $M_Q$,
and further estimate $DAG(Q)$. The exponential bounds for the upper and lower tails of the distribution of $M_Q$ were discussed in \cite{janson1990poisson,janson2004upper}. 
\end{IEEEproof}

\section{Proof of Theorem 2}
\label{app2}
\begin{IEEEproof}
In Chung-Lu model \cite{chung2002connected}, given $n$ nodes and a degree sequence $(d_1,d_2,\ldots,d_n)$,
an edge is added between any two nodes $u$ and $v$ with the probability $p_{u,v}=\frac{d_ud_v}{\sum_{k\leq n}d_k}.$
It is assumed that $\max\{d_k^2\}\leq \sum_{k\leq n} d_k$ \cite{chung2002connected}.
Besides, the relation of $n$ and $\alpha,\beta$ is 
\begin{equation}
\label{n-alpha-beta}
n=\sum_{d=1}^\infty n(d)=\sum_{d=1}^\infty \alpha d^{-\beta} = \frac{\alpha}{\beta-1}.
\end{equation}

For the case of exact and partial knowledge with attribute ignored, the expected number of matches yielded 
from querying $G$ with $Q$ can be computed as follows. 
Let $I$ be any induced subgraph of size $n_Q$ of $G$ with the nodes mapped to nodes in $Q$, \eg, $u',v'$ in $I$ correspond to
$u,v$ in $Q$.
\begin{equation}
\label{M_Q_power}
\begin{aligned}
M_Q=& \mathbb{E}(\sum_I\prod_{(u,v)\in E_Q}p_{u',v'}) \\
=& \mathbb{E}(\sum_I\prod_{(u,v)\in E_Q}\frac{d_{u'}d_{v'}}{\sum_{k\leq n}d_k}) \\
=& \frac{1}{(\sum_{k\leq n}d_k)^{m_Q}}\sum_I \mathbb{E}(\prod_{(u,v)\in E_Q}d_{u'}d_{v'}) 
\end{aligned}
\end{equation}

For any two edges $(u'_1,v'_1),(u'_2,v'_2)$ in $I$, they might be adjacent, \eg $u'_1=u'_2$.
So $d_{u'_1}d_{v'_1}$ and $d_{u'_2}d_{v'_2}$ are either independent or positively correlated.
Thus we have $cov(d_{u'_1}d_{v'_1}, d_{u'_2}d_{v'_2})\geq 0$, and then
\begin{equation}
\begin{aligned}
& \mathbb{E}(d_{u'_1}d_{v'_1}\cdot d_{u'_2}d_{v'_2}) \\
=& cov(d_{u'_1}d_{v'_1}, d_{u'_2}d_{v'_2})+\mathbb{E}(d_{u'_1}d_{v'_1}) \mathbb{E}(d_{u'_2}d_{v'_2}) \\
\geq & \mathbb{E}(d_{u'_1}d_{v'_1}) \mathbb{E}(d_{u'_2}d_{v'_2}) \\
= & \mathbb{E}(d_{u'_1}) \mathbb{E}(d_{v'_1}) \mathbb{E}(d_{u'_2})\mathbb{E}(d_{v'_2}) \\
=& \bar{d}^4,
\end{aligned}
\end{equation}
where $\bar{d}$ is the average degree.
Likewise, we have
\begin{equation}
\mathbb{E}(\prod_{(u,v)\in E_Q}d_{u'}d_{v'} )\geq \bar{d}^{2m_Q}.
\end{equation}
The sum of degrees in $G$ is
\begin{equation}
\begin{aligned}
\sum_{k\leq n}d_k= & \sum_{d=1}^{\infty}\alpha d^{-\beta}\cdot d \\
=& \alpha\sum_{d=1}^{\infty}d^{1-\beta} \\
=& \frac{\alpha}{\beta-2}
\end{aligned}
\end{equation}
so the average degree is 
\begin{equation}
\bar{d}=\frac{1}{n}\sum_{k\leq n}d_k = \frac{\alpha}{(\beta-2)n}.
\end{equation}

By Equation (\ref{M_Q_power}), we have
\begin{equation}
\begin{aligned}
M_Q \geq & (\frac{\beta-2}{\alpha})^{m_Q}\sum_I d^{2m_Q} \\
\geq & (\frac{\beta-2}{\alpha})^{m_Q}\sum_I (\frac{\alpha}{(\beta-2)n})^{2m_Q} \\
=& (\frac{\beta-2}{\alpha})^{m_Q}\cdot \binom{n}{n_Q}n_Q!(\frac{\alpha}{(\beta-2)n})^{2m_Q} \\
=& \binom{n}{n_Q}n_Q!(\frac{\alpha}{(\beta-2)n^2})^{m_Q}.
\end{aligned}
\end{equation}
We will obtain the lower bound after substituting $\alpha$ with $n,\beta$ by equation \ref{n-alpha-beta}.
\end{IEEEproof}


{\fontsize{9pt}{1em}
{\bibliographystyle{ieee}
\bibliography{moreKnowledge}
}}

\end{document}